\documentclass[aps, prd, 10pt,notitlepage,nofootinbib,superscriptaddress,twocolumn,groupedaddress]{revtex4}

\usepackage{graphicx,color}
\usepackage{amstext,amsmath,amssymb,amsfonts,bbm}
\usepackage{amsmath,amsthm}
\usepackage{amssymb}
\usepackage{graphicx}

\newcommand{\Ref}[1]{(\ref{#1})}

\def\f{\frac}

\newcommand{\bra}[1]{\langle {#1}|}
\newcommand{\ket}[1]{| #1 \rangle}

\newtheorem{Theorem}{Theorem}[section]

\newcommand{\startproof}{\textbf{Proof:\ \ }}
\newcommand{\finishproof}{\hfill $\Box$ \\}

\newcommand{\R}{\mathbb{R}}

\newcommand{\h}{{\cal H}}

\newcommand{\eqa}{\begin{eqnarray}}
\newcommand{\neqa}{\end{eqnarray}}
\newcommand{\be}{\begin{equation}}
\newcommand{\ee}{\end{equation}}

\newcommand{\bn}{\vec{n}_f}
\newcommand{\bl}{|\Sigma_{f}|}
\newcommand{\bv}{\vec{\Sigma}_{f}}

\newcommand{\slc}{SL(2,\mathbb{C})}

\def\be{\begin{eqnarray}}
\def\ee{\end{eqnarray}}

\newcommand{\ch}{\mathcal H}
\newcommand{\ci}{\mathcal I}

\newcommand{\ck}{\mathcal K}

\newcommand{\cs}{\mathcal S}

       \newcommand{\Fraki}{\mathfrak{I}}

\renewcommand{\a}{\alpha}

\newcommand{\g}{\gamma}

\newcommand{\eps}{\epsilon}

\newcommand{\sig}{\sigma}
\newcommand{\Sig}{\cal S}

\renewcommand{\t}{e}

\newcommand{\rmd}{\mathrm d}

\newcommand{\lt}{\left}
\newcommand{\rt}{\right}

\newcommand{\lag}{\left\langle}
\newcommand{\rag}{\right\rangle}

\newcommand{\tr}{\mathrm{tr}}
\newcommand{\bbc}{\mathbb{C}}

\begin{document}

\title{\bf Generalized Spinfoams}

\author{You Ding}
\affiliation{Centre de Physique Th\'eorique%
 \footnote{Unit\'e mixte de recherche (UMR 6207) du CNRS et des Universit\'es de Provence (Aix-Marseille I), de la Meditarran\'ee (Aix-Marseille II) et du Sud (Toulon-Var); laboratoire affili\'e \`a la FRUMAM (FR 2291).}, CNRS-Luminy Case 907,  F-13288 Marseille, EU}
\author{Muxin Han}
\affiliation{Centre de Physique Th\'eorique%
 \footnote{Unit\'e mixte de recherche (UMR 6207) du CNRS et des Universit\'es de Provence (Aix-Marseille I), de la Meditarran\'ee (Aix-Marseille II) et du Sud (Toulon-Var); laboratoire affili\'e \`a la FRUMAM (FR 2291).}, CNRS-Luminy Case 907,  F-13288 Marseille, EU}
\affiliation{Max-Planck-Institut f\"ur Gravitationsphysik, Am M\"uhlenberg 1, D-14476 Golm, EU}
\author{Carlo Rovelli}
\affiliation{Centre de Physique Th\'eorique%
 \footnote{Unit\'e mixte de recherche (UMR 6207) du CNRS et des Universit\'es de Provence (Aix-Marseille I), de la Meditarran\'ee (Aix-Marseille II) et du Sud (Toulon-Var); laboratoire affili\'e \`a la FRUMAM (FR 2291).}, CNRS-Luminy Case 907,  F-13288 Marseille, EU}

\date{\small\today}

\begin{abstract}
\noindent
We reconsider the spinfoam dynamics that has been recently introduced, in the generalized Kami\'nski-Kisielowski-Lewandowski (KKL) version where the foam is not dual to a triangulation.  We study the Euclidean as well as the Lorentzian case.  We show that this theory can still be obtained as a constrained BF theory satisfying the simplicity constraint, now discretized on a general oriented 2-cell complex.  This  constraint implies that boundary states admit a (quantum) geometrical interpretation in terms of polyhedra, generalizing the tetrahedral geometry of the simplicial case. We also point out that the general solution to this constraint (imposed weakly) depends on a quantum number $r_f$ in addition to those of loop quantum gravity. We compute the vertex amplitude and recover the KKL amplitude in the Euclidean theory when $r_f\!=\!0$. We comment on the eventual physical relevance of $r_f$, and the formal way to eliminate it.
\end{abstract}

\maketitle




\section{Introduction}\label{Introduction}

\noindent
The spinfoam formalism  \cite{simple,sfrevs,BC} offers a formulation of the dynamics of quantum gravity strictly related to loop quantum gravity (LQG)\cite{thiemannbook,rovellibook,rev}.  The precise relation between the two approaches is well-understood in 3 dimensions \cite{perez}, and under study in 4 dimensions \cite{links}.

The spinfoam theory introduced in \cite{EPRL,FK} can be derived starting from the Plebanski formulation of GR \cite{plebanski} (including the Barbero-Immirzi
 parameter $\g$), and defined as a BF theory discretized on a simplicial cellular complex and constrained by the so called simplicity constraint.  The constraint can be imposed using the master-constraint technique \cite{QSD,EPRL}, or, more simply,  using the Gupta--Bleuler procedure, namely asking the matrix elements of the constraint to vanish on physical states \cite{DingYou}.  The resulting model has remarkable properties: (i) the boundary states have a geometrical interpretation in terms of quantum tetrahedral geometry \cite{barbieri,BC}; (ii)  there are strong indications that the semiclassical behavior of the theory matches classical general relativity \cite{semiclassical,semiclassical2,FC,Bianchi:2010zs}, thus correcting difficulties of earlier models \cite{BCtrouble}; and (iii) the boundary kinematics is strictly related to that of LQG \cite{EPRL,DingYou}.

The relation with LQG, however, is limited by the fact that the simplicial-spinfoam boundary states include only four-valent spin networks.  This is a drastic reduction of the LQG state space.
In \cite{kkl}, Kami\'nski, Kisielowski, and Lewandowski (KKL) have considered a generalization of the spinfoam formalism to spin networks of arbitrary valence, and have constructed a corresponding  vertex amplitude. This generalization provides truncated transition amplitudes between \emph{any} two LQG states \cite{simple}, thus correcting the limitation of the relation between the model and LQG. This generalization, on the other hand, gives rise to several questions. The KKL vertex is obtained via a ``natural" mathematical generalization of the simplicial Euclidean vertex amplitude.  Is the resulting vertex amplitude still related to constrained BF theory (and therefore to GR)? In particular, do KKL states satisfy the simplicity constraint?   Can we associate to these states a geometrical interpretation similar to the one of the simplicial case?  Can the construction be extended to the physically relevant Lorentzian case?

Here we answer several of these questions. We show that it is possible to start form a discretization of BF theory on a general 2-cell complex, and impose the same boundary constraints that one impose in the simplicial case (simplicity and closure).   Remarkably,  on the one hand, they reduce the BF vertex amplitude to a (generalization of) the KKL vertex amplitude, in the Euclidean case studied by KKL.  On the other hand, a theorem by Minkowski \cite{Minkowski} garantees that these constraints are precisely those needed to equip the classical limit of each truncation of the boundary state space to a finite graph, with a geometrical interpretation, which turns out to be in terms of polyedra \cite{poly}.

These results reinforce the overall coherence of the generalized spinfoam formalism.

Surprisingly, however, the state space defined by imposing the simplicity constraint weakly is larger than the one of quantum gravity.  It includes one additional degree of freedom, described by a new quantum number $r_f$.\footnote{The enlargement is not an effect from the generalization to arbitrary 2-cell complexes. The Hilbert space is enlarged also in the simplicial case, compared with the state space defined in \cite{EPRL}. This additional quantum number was first noticed by Sergei Alexandrov \cite{alexei}.} The quantum number $r_f$ affects non-trivially both the face amplitude and the vertex amplitude of the model.   The quantum number $r_f$ is frozen if in addition to the weak imposition of the (linear) simplicity constraint, we also impose strongly a diagonal quadratic constraint. With a suitable operator ordering of this constraint, the state space can be reduced back down to the LQG state space.

Does the $r_f$ quantum number have physical relevance? If we take the principle that the quantum theory we are seeking has the same number of degrees of freedom as the classical theory, then the answer is negative. This principle indicates that the appropriate way of imposing the constraints is the one that gets rids of the extra states.   However, we think it is nevertheless interesting to keep in mind the existence of these additional solutions to the weak simplicity constraints. We comment more on this in the conclusion.

An outline for the article is as follows. In Section \ref{spinfoam Representation of BF Theory}, we review the spinfoam representation of the BF partition function on a general complex, and we discuss the structure of the boundary Hilbert space of BF theory. In Section \ref{Implementation of Boundary Quantum Geometry}, we implement the geometric constraint to the BF boundary Hilbert space. After solving the constraint weakly, two new boundary Hilbert space are constructed for both the Euclidean and the Lorentzian theory. We also show that the new boundary Hilbert space carries a representation of quantum polyhedral geometry. In Section \ref{New spinfoam Model}, we derive the new spinfoam vertex amplitude and face amplitude from the new boundary Hilbert space. In Section \ref{conclusion}, we conclude and point out the open issues. We assume that the Barbero-Immirzi parameter $\gamma$ is positive.

\begin{figure}[h]
\begin{center}
\includegraphics[width=7cm]{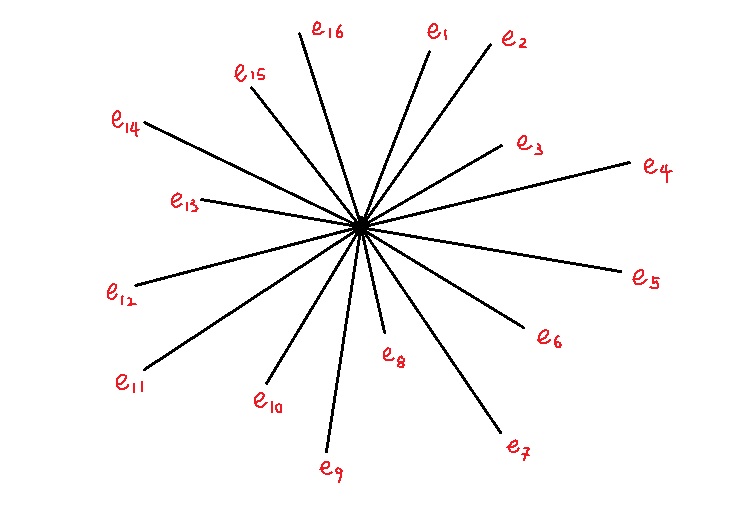}
\caption{A generalized spinfoam vertex.}
\label{complex}
\end{center}
\end{figure}


\section{spinfoam Representation of BF Theory}\label{spinfoam Representation of BF Theory}

We start with a brief review of the construction of the BF spinfoam partition function and the structure of its boundary Hilbert space \cite{Ooguri}, which is the starting point of the definition of the theory.  The BF partition function is formally defined by the path integral
\be
Z_{BF}:=\int\  DA \; DB \ \ \exp \big(i\int_{M} \tr(B\wedge F[A])\big)
\label{fpi}
\ee
where $B$ is a 2-form field  on the manifold $M$,  with values in the Lie algebra $\mathfrak{g}$ of a group $G$ and $F$ is the curvature of the $G$-connection $A$. Here we take the internal gauge group $G$ to be either $G=Spin(4)$ (for the Euclidean case) or $G=SL(2,\mathbb{C})$ (for the Lorentzian case). A formal integration over $B$ gives
\be
Z_{BF}=\int  DA \prod_{x\in M}\delta( F[A]) \label{BF}
\ee
which is an integration over the flat connections. In order to make sense of the formal path integral (\ref{BF}), we discretize it. However, instead of discretizing the path integral on an oriented 2-complex dual to a simplicial decomposition of the manifold $M$ as is usually done, we introduce here an arbitrary oriented 2-complex $\ck$ (as in \cite{kkl}) with or without boundary.

We take a combinatorial definition of an oriented 2-complex.  An oriented 2-complex $\ck:=(V(\ck),E(\ck),F(\ck)$ consists of sets of \emph{vertices} $v\in V(\ck)$, \emph{edges} $e\in E(\ck)$ and \emph{faces} $f\in F(\ck)$, equipped with a \emph{boundary relation} $\partial$ associating an ordered pair of vertices $(s(e), t(e))$ (``source" and ``target") to each edge $e$ and a finite sequence of edges $\{e_k^{\epsilon_{e_{k}f}}\}_{k=1,...,n}$ to each face $f$, with $t(e_k)=s(e_{k+1})$, $t(e_n)=s(e_1)$ and  $\epsilon_{ef}=\pm1$; here we call $e^{-1}$ the edge with reversed order of $e$.
We let $\partial  f$ denote the cyclically ordered set of edges that bound the face $f$, or (if it is clear from the context) the cyclically ordered set of vertices that bound the boundary edges of $f$. We also write $\partial  v$ to indicate the set of edges bounded by $v$, and of faces that have $v$ in their boundary.  Similarly, we write $\partial e$ to indicate the set of the faces bounded by $e$.  When $e\in\partial f$, we define $\epsilon_{ef}=1$ if the orientation of $e$ is consistent with the one induced by the face $f$ and  $\epsilon_{ef}=-1$ if it is not.

The boundary graph $\g=\partial \ck$ is a 1-cell subcomplex of $\ck$. An edge $e\in E(\ck)$ is an edge of the boundary graph $\g$ if and only if it is contained in only one face, otherwise it is an internal edge. A vertex $v\in V(\ck)$ is a vertex of the boundary graph $\g$ if and only if it is contained in exactly one internal edge of $\ck$, otherwise it is an internal vertex of $\ck$. We assume boundary vertices and boundary edges to form a graph, which is the boundary of the two-complex.

We introduce also the notion of the boundary graph $\gamma_v$ of a single vertex $v$. This is the graph whose nodes are the edges $e$ in $\partial v$ and whose links are the faces $f$ in $\partial v$. The boundary relation defining the graph is the relation $e\in\partial f$ and the orientation of the links is the one induced by the faces. The graph $\gamma_v$ can be visualized as the intersection between the two complex and a small sphere surrounding the vertex.

\begin{figure}[h]
\begin{center}\vskip3mm
\includegraphics[width=6cm]{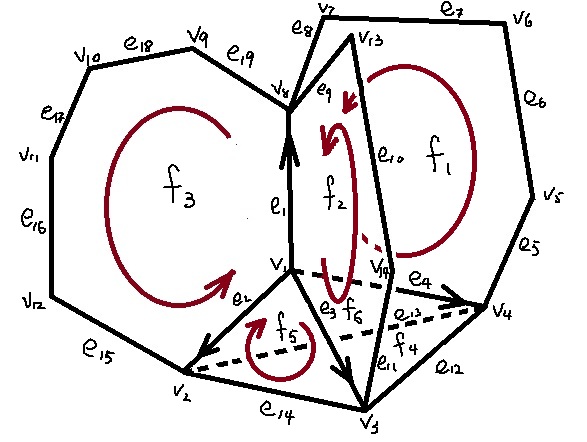}
\caption{An oriented 2-cell complex $\ck:=(F(\ck),E(\ck),V(\ck))$, where $F(\ck)=\{f_1,\cdots,f_6\}$, $E(\ck)=\{e_1,\cdots,e_{19}\}$, $V(\ck)=\{v_1,\cdots,v_{14}\}$. $v_1$ is internal vertex, and $e_1,e_2,e_3,e_4$ are internal edges, while all other edges and vertices belong to the boundary graph $\g=\partial \ck$.}
\label{complex}
\end{center}
\end{figure}

We discretize the BF partition function on the oriented 2-cell complex $\ck$, by replacing the continuous field $A$ with the assignment of an element of $G$ to each edge. By convention, $g_{e^{-1}}:=g^{-1}_{e}$. Then equation \eqref{BF} becomes
\be
Z_{BF}(\ck)=\int  dg_e\  \prod_{f}\ \delta\big(\!\!\prod_{e\in \partial  f} g_e^{\epsilon_{ef}}\big) \label{BFd},
\label{zbf}
\ee
where $dg_e$ is the product over all the edges of the Haar measure, the product over $f$ is over all the faces of $\ck$ and the product over $e$ is the product over the edges bounding the face $f$ of the group element associated
 to these edges, ordered by the orientation of the face.  This is the partition function of BF theory.

We now express this partition function as a sum over representations and intertwiners. For this, it is convenient to treat the Euclidean and Lorentzian cases separately.

\subsection{Spin(4) BF Theory}\label{Spin(4) BF Theory}

Consider the Euclidean case $G=Spin(4)$. The delta function on $Spin(4)$ can be expanded in irreducible representations
\be
\delta(U)=\sum_{\rho}\dim(\rho)\chi^\rho(U)\label{delta0}
\ee
where $\rho=(j^+,j^-)$ labels the unitary irrep of Spin(4), $\dim(\rho)=(2j^++1)(2j^-+1)$ is the dimension of the representation space, and $\chi_\rho$ is the character of the representation $\rho$. Irreducible representations can also be conveniently labelled with the two half integers $k=j^++j^-$ and $p=j^+-j^-.$

Expanding the delta function in representations,  (\ref{BF}) becomes
\be
Z_{BF}(\ck)&=&\int dg_e \ \prod_f \left(\sum_{\rho}
\dim(\rho)\ \chi^{\rho}(U_f)\right)
\nonumber \\
&=&\sum_{\rho_f}\int dg_e\  \prod_f \dim(\rho_f)\ \chi^{\rho_f}(U_f)\label{BF1}.
\ee
This is the expression for the spinfoam amplitude in the group element basis. Let us now translate this into the more common representations-intertwiners basis.

This can be obtained by performing the integrals, precisely as in the simplicial case. We have one integration per edge, of the form
\be
K_{\mathbf{M},\mathbf{N}}=\int dg_{e} \prod_{f\in \partial e}
\Pi^{\rho_{f}}_{M_{f}N_{f}}\!(g^{\epsilon_{ef}}_{e})
\label{int}
\ee
where $\Pi^\rho_{MN}\!(g)$ is the matrix element of the Spin(4) representation $\rho$;  $\mathbf{M}=M_{f_1},...,M_{f_n}$ is a multi-index;  and the product is over the $n$ faces bounded by $e$ (including repeated faces). In the case where $\ck$ is dual to a simplicial complex, $n\!=\!4$.
It is immediate to see that $K_{\mathbf{M},\mathbf{N}}$ is the operator in
the tensor product
$(\bigotimes_{f_{out}}\rho_f)\otimes(\bigotimes_{f_{in}}\ \rho_f^\dagger)$
of the $\rho_f$ representation spaces
(where $f_{in}$ are the faces with the same orientation as $e$ and $f_{out}$ are the faces
with opposite orientation.) that projects on its invariant subspace
\be
{\cal H}_e=
\mathrm{Inv}\big[(\bigotimes_{f_{out}}\rho_f)\otimes(\bigotimes_{f_{in}}\ \rho_f^\dagger)
\big].
\ee
Let $I$ label an orthonormal basis in ${\cal H}_e$. (These are called intertwiners.) Then
\be
K_{\mathbf{M},\mathbf{N}}=\sum_{I}\  I_{\mathbf{M}}\  I^\dagger_{\mathbf{N}}.
\ee
For each internal edge $e$, the two intertwiners are associated to the two vertices bounding the edge (see Figure 3), in the sense that their indices are contracted with the other intertwiners at the same vertex.
\begin{figure}[h]
\begin{center}\vskip3mm
\includegraphics[width=5cm]{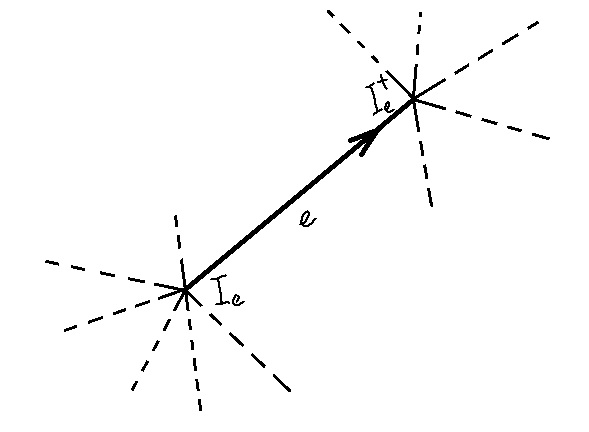}
\caption{Assign $I_e$ to the begin point and assign $I^\dagger_e$ to the end point of an internal edge $e$.}
\label{edge}
\end{center}
\end{figure}
The result of the integration is therefore
  \be
  Z_{BF}(\ck)=\sum_{\rho_f}\prod_{f}\dim(\rho_f)\sum_{I_e}\prod_{v}A_v(\rho_f,I_e).\label{BF2}
  \ee
Here the sum over $I_e$ is over the assignment of one intertwiner to each edge $e$ of $\ck$. The product over $v$ is over the vertices of $\ck$. The vertex amplitude $A_v(\rho_f,I_e)$ is defined as follows.
Say at the vertex $v\in V(\ck)$ there are $n$ outgoing edges $e_{out}$ and $m$ incoming edges $e_{in}$. Then
      \be
      A_v(\rho_f,I_e):=\tr\lt(\bigotimes_{e_{out}}I_{e_{out}}\bigotimes_{e_{in}}I_{e_{in}}^\dagger\rt)
      \label{trace}
      \ee
The trace in eq.(\ref{trace}) is precisely the spinfoam trace defined in \cite{kkl}. The contractions between the intertwiners in the spinfoam trace could be described by the follows: For each edge $e$ each index $M_i$ is associated with a face $f$ bounded by the edge $e$. The trace is defined by contracting the two indices associated with the same face of the two intertwiners corresponding to the two edges bounding $f$. This can be easily seen to give the character $\chi^\rho$ of (\ref{BF1}). In the special case when the complex $\ck$ is dual to a simplicial complex, there are 5 internal edges joining at $v$ and each pair of edges determines a 2-face, the spinfoam trace is nothing but the Spin(4) 15-j symbol.

Alternatively, the BF partition function can also be expressed in the form \cite{kkl}
  \be
  Z_{BF}(\ck)=\sum_{\rho_f}\prod_{f}\dim(\rho_f)\ \tr\lt(\bigotimes_{e\in E(\ck)}P_e\rt)
  \ee
  where $P_e:=\sum_{I_e}I_e\otimes I_e^\dagger$ is understood as the projection operator projecting from the product of the representations on the 2-faces bounded by $e$ to its invariant subspace. And the index contractions in $\tr\lt(\otimes_{e\in E(\ck)}P_e\rt)$ are the contractions between intertwiners, as above.

All gravitational spinfoam theories have this same structure.

\subsection{SL(2,$\bbc$) BF Theory}\label{SL2C BF Theory}

Let now $G=SL(2,\bbc)$. The derivation of the spinfoam representation of $SL(2,\bbc)$ is as above, with a few differences.  $SL(2,\bbc)$ unitary irreps  (in the principle series) can be labelled by the same quantum numbers $(k,p)$ as the $SO(4)$ ones, but now $p$ is a real number \cite{ruhl}.  The unitary irreps of $SL(2,\bbc)$ are infinite dimensional and can be decomposed into an infinite direct sum of SU(2) irreps, i.e.
\be
V^{(k,p)}=\bigoplus_{j=k}^\infty V_j^{(k,p)}
\ee
where $V_j^{(k,p)}\sim V_j$ is the carrier space of the spin$j$ representation of SU(2).
This decomposition provides a convenient basis $|j,m>$ in $V^{(k,p)}$, obtained diagonalizing $L^2$ and $L^z$ of SU(2). In this basis, for $g\in SL(2,\bbc)$, we write the representation matrices on $V^{(k,p)}$ as $\Pi^{(k,p)}_{jm,j'm'}(g)$ where $j\in\{k,k+1,\cdots,\infty\}$ and $m\in\{-j,\cdots,j\}$. As one might expect from the fact that $p$ is a continuous label, the representation ``matrix element'' $\Pi^{(k,p)}_{jm,j'm'}$ is distributional on the Hilbert space $L^2[SL(2,\bbc)]$ defined by the Haar measure. These matrix elements form a generalized orthonormal basis and define a Fourier-like transform. That is, for any square integrable function $f(g)$ on $\slc$,
\be
&&f(g)=\frac{1}{8\pi^4}\sum_k\int_{-\infty}^{+\infty}\!\!\!\!\!\!\rmd p\, (k^2+p^2)\,  \tr\!\lt[F(k,p)\, \Pi^{(k,p)}(g^{-1})\rt] \nonumber\\
&&F(k,p)=\int_{\slc} f(g)\ \Pi^{(k,p)}(g)\ \rmd\mu_H(g)
\ee
which is known as Plancherel theorem \cite{ruhl}. Accordingly, we have an identity for Fourier decomposition of delta function on $\slc$
\be
\delta(g)=\frac{1}{8\pi^4}\sum_k\int_{-\infty}^{+\infty}\tr\lt[\Pi^{(k,p)}(g)\rt](k^2+p^2)\ \rmd p
\ee
in analogy with eq.(\ref{delta0}). Proceeding as in the Euclidean case, we find
\be
Z_{BF}(\ck)&=&\int\prod_{e}\rmd g_{e}\ \prod_{f}\delta(U_f)\\ \nonumber
&&\hspace{-5em}= \ \  \sum_{k_f}\int \rmd p_f\prod_{f}(k_f^2+p_f^2)\int\rmd g_{e}\prod_{f}\tr\lt[\Pi^{(k_f,p_f)}(U_f)\rt]\label{BF3}
\ee
As in the euclidean case, each $g_e$ integral is of the form
\be
K_{\mathbf{j}\mathbf{m},\mathbf{j'}\mathbf{m'}}&=&\int {\rmd g_e} \prod_{f\in\partial  e} \Pi^{(k_{f},p_{f})}_{j_{f}m_{f},j'_{f}m'_{f}}\lt(g^{\epsilon_{ef}}_{e}\rt).
\label{int3}
\ee
Formally, this is still a projector on the invariant component of the tensor product of $n$ irreducibles.  However, since now one of the two Casimirs has continuous spectrum $p$, then the trivial representation $p=k=0$ is not a proper subspace of the tensor product, but only a generalized subspace.  This does not forbids us to introduce an orthonormal basis of intertwiners $I$ in this subspace, as we did in the Euclidean case, and write
\be
K_{\mathbf{j}\mathbf{m},\mathbf{j'}\mathbf{m'}}&=&\sum_I\ I_{\mathbf{j}\mathbf{m}} \ I^\dagger_{\mathbf{j'}\mathbf{m'}}
\ee
but we have to remember that the intertwiners are generalized vectors.
Using this, we can formulate the spinfoam representation of $\slc$ BF theory in the same way as we did for Spin(4) theory.
\begin{itemize}
\item The Fourier decomposition of the $\slc$ delta function assigns an $\slc$ irrep labeled by $(k_f,p_f)$  to each face $f$.

  \item Eq.(\ref{int3}) assigns an $\slc$ intertwiner $I^e$ to each source of each edge $e$, and a dual intertwiner $I^{e\dagger}$ to its target.

  \item At each vertex $v$ with $n$ outgoing edges $e^{out}_1,\cdots,e^{out}_n$ and $m$ incoming edges $e^{in}_1,\cdots,e^{in}_m$, the intertwiners $I^{e^{out}}$ and $I^{e^{in}\dagger}$ are contracting on their $\mathbf{j},\mathbf{m}$ and $\mathbf{j}',\mathbf{m}'$ indices, according to how the faces neighboring the vertex are bounded by the edges. The result of this contraction gives the spinfoam vertex amplitude
        \be
      A_v\Big((k,p)_f,I_e\Big):=\tr\lt(\left(\bigotimes_{e_{out}}I_e\right)\otimes\left(\bigotimes_{e_{in}} I_e^\dagger\right)\rt)\label{vertex}
      \ee
  \item Finally the partition function of $\slc$ BF theory is
  \be
  Z_{BF}=\sum_{k_f I_e}\int \rmd p_f\prod_{f}(k_f^2+p_f^2)\prod_{v}A_v\Big((k,p)_f,I_e\Big)
  \label{vertex2}
  \ee
\end{itemize}
This expression, however, is ill defined, due to the fact that the intertwiners are generalized vectors, and the trace \eqref{vertex} may diverge.
This issue is addressed and answered in \cite{finite}, where it is shown that
the source o f the divergence is a redundant integral over $\slc$ in the definition of $A_v$. It is then immediate to regularize $A_v$ by removing one  $\slc$ integration per each vertex. The resulting amplitude is proven in  \cite{finite} to be finite, except for some particular pathological vertices, which we exclude here for simplicity.  In what follows we always assume that the vertex amplitude is so renormalized.

\subsection{Boundary Hilbert Space}\label{Boundary Hilbert Space}

Let us rewrite the partition function \eqref{BFd} in a slightly different form. Split each edge $e$ bounded by the vertices $v$ and $v'$ into two half edges $(ev)$ and $(ev')$, and associate a group element $g_{ev}$
 to each half edge (oriented towards the vertex). Then replace each integral $dg_e$ with the two integrals $dg_{ev}$, $dg_{ev'}$. This gives
\be
Z_{BF}(\ck)=\int  dg_{ev}\  \prod_{f} \delta\big(\!\!\prod_{e\in \partial  g} (g^{-1}_{ev}g_{ev'})^{\epsilon_{ef}}\big) \label{BFd2},
\label{zbf2}
\ee
where there is one integration per each couple vertex/adjacent-edge. Next, let $v$ be a vertex in the boundary of the face $f$. For each such couple $f\!v$, introduce a group variable $g_{\!f\!v}$. Then \eqref{zbf2} can be rewritten in the form
\be
Z_{BF}(\ck)\!=\!\!\int dg_{\!f\!v} dg_{ev} \prod_{f} \delta(\!\!\prod_{v\in\partial  f}\!\!g_{f\!v}) \prod_{f\!v}\delta(g^{-1}_{\!f\!v} g_{ev}g^{-1}_{e'v})
\label{BFd3}
\label{zbf3}
\ee
where $e$ and $e'$ are the two edges in the boundary of $f$ that meet at $v$, ordered by the orientation of $f$. This can be rewritten in the form
\be
Z_{BF}(\ck)=\int dg_{\!f\!v} \ \prod_{f}\  \delta \big(\!\!\prod_{v\in\partial  f}\!\!g_{\!f\!v}\big) \ \prod_{v}A_v(g_{\!f\!v})\label{hr}
\ee
where the vertex amplitude $A_v(g_{f})$ is defined by
\be
A_v(g_f)=\int \prod_{e\in \partial  v} dg_{e}\ \
\prod_{f\in\partial  v}\delta(g_{e_f}g_fg^{-1}_{e'_f})
\label{verterxa}
\ee
is a function of one group element for each face in the boundary of $v$.  Here the integral is over one group element per each edge in the boundary of the vertex $v$ and, as before, $e$ and $e'$ are the two edges in the boundary of $f$ that meet at $v$. This is the ``holonomy" form of the partition function \cite{Magliaro:2010ih}.

Let $|F_v|$ be the number of links $f$ of the graph $\gamma_v$, namely the number of faces $f$ in $\partial v$. The vertex amplitude \eqref{verterxa} is a function in
\be
\ch_{\g_v}=L_2[G^{|F_\gamma|}].
\ee
We call this the (non-gauge invariant) boundary Hilbert space of the vertex $v$.  It is easy to se that the vertex amplitude \eqref{verterxa} is an element of this space. More precisely, it is an element of the (possibly generalized) subspace
\be
{\cal K}_{\g_v}=L_2[G^{|F_\gamma|}/G^{|E_\gamma|}]
\ee
where $|E_\gamma|$ is the number of nodes of $\gamma_v$, namely the number of edges in $\partial v$, formed by the states invariant the gauge transformation
\be
\psi(g_e)=\psi(\Lambda_{s_e}g_e\Lambda_{t_e})\label{gauge}
\ee
where $\Lambda\in G$ and $s_e$ and $t_e$ are the source and target of $e$.

A moment of reflection shows also that \eqref{trace} and \eqref{vertex} are simply the amplitude
\eqref{verterxa} expressed in the standard spin network basis of ${\cal K}_{\g_v}$. Let us now study the boundary space $\ch_{\g_v}$ in more detail.
(It is convenient to consider the non-gauge-invariant Hilbert space $\ch_{\g_v}$, besides the gauge invariant one because the expressions of geometric constraints will not be gauge invariant, thus they can only be represented as operators on $\ch_{\g_v}$.)

The natural derivative operator defined on the Hilbert space $L_2[G]$ is the left invariant derivative that generates the right $G$ action:
\be
{J}^{IJ}\psi(g)&=&\frac{\rmd}{\rmd \a}\psi(e^{\a T^{IJ}}g)\Big|_{\a=0}
\ee
where $T^{IJ}$ ($I,J=0,\cdots,3$) is a standard Lie algebra generator of $Lie(G)$.
%

Fix an SU(2) subgroup of $G$, and choose a basis in $Lie(G)$ such that the direction $I=0$ is preserved by $SU(2)$. Then we can split the six generators $T^{IJ}$ of $Lie(G)$ into 3 rotation generators and 3 boost generators. Accordingly, we define ($i,j,k=1,2,3$)
\be
{L}^i:=\frac{1}{2}\eps^{i}_{\ jk}{J}^{jk},\ \ \ \ \ \ {K}^i:={J}^{0i}
\ee
which have the standard commutation relations
\be
\lt[{L}^i,{L}^j\rt]&=&\eps^{ij}_{\ \ k}{L}^k, \\ \lt[{K}^i,{K}^j\rt]&=&s\eps^{ij}_{\ \ k}{L}^k,\\  \lt[{K}^i,{L}^j\rt]&=&\eps^{ij}_{\ \ k}{K}^k
\ee
where $s=+1$ for $Spin(4)$ and $s=-1$ for $\slc$.


We denote by ${J}_f^{IJ}$ the left invariant derivative operator acting on the variable $g_f$ of $\psi(g_f)\in\ch_{\g_v}$.
Notice that the right invariant vector field
\be
{R}^{IJ}\psi(g)&=&\frac{\rmd}{\rmd \a}\psi(g e^{\a T^{IJ}})\Big|_{\a=0}
\ee
satisfies ${R}^{IJ}\psi(g)={J}^{IJ}\psi(g^{-1})$. Therefore
\be
J_{f^{-1}}^{IJ}={R}_f^{IJ}.
\ee
The bivector operators $J_f^{IJ}$ have a physical interpretation in terms of the $BF$ theory we started from. They are the quantum operators that quantize the discretized version of the 2-form field $B^{IJ}$, restricted to a 3-dimensional boundary. The reason for this is the follows: Classically the Hamiltonian analysis of BF theory can be carried out \cite{BHNR}. The resulting non-vanishing Poisson bracket reads
\be
\Big\{\eps^{abc}B_{abIJ}(x),A^{KL}_d(x')\Big\}=\delta^c_d\delta^K_{[I}\delta^L_{J]}\delta^3(x,x')
\ee
where $a,b,c=1,2,3$, $x$ and $x'$ belong to a 3-dimensional spatial manifold $\Sig$. These canonical conjugate variables can be discretized in analogy with Hamiltonian lattice gauge theory. Given a graph $\g$ imbedded in $\Sig$, there exists a 2-cell complex dual to the graph $\g$, such that given a link $f$ in the graph there is a unique 2-face $S_f$ dual to the link $f$. This 2-cell complex defines a polyhedral decomposition of the spatial manifold $\sig$. With this setting, we associate a group variable $g_f\in G$ to each link $f$, and associate a Lie algebra variable $B_f^{IJ}$ to each $S_f$ (the Lie algebra variables are also labeled by $f$ because of the 1-to-1 correspondence between links and 2-faces). The Poisson algebra of these discretized variables has the following standard expression
\be
\Big\{g_f,g_{f'}\Big\}&=&0\nonumber\\
\Big\{B^{IJ}_f,g_{f'}\Big\}&=&\delta_{ff'}T^{IJ}g_f\nonumber\\
\Big\{B_f^{IJ},B^{KL}_{f'}\Big\}&=&\delta_{ff'}f^{IJ,KL}_{\ \ \ \ \ \ MN}B^{MN}_f
\ee
where $f^{IJ,KL}_{\ \ \ \ \ \ MN}$ denotes the structure constant of $Lie(G)$. In our case, if we consider our boundary graph $\g_v$ and abstractly define the above Poisson algebra on $\g_v$, we find that the bivector operator ${J}^{IJ}_f$ for each oriented link $f$ (as a right invariant vector) is the quantum operator representing the Lie algebra variable $B_f^{IJ}$ (up to $-i\hbar$), because of the commutation relation between ${J}_f^{IJ}$ and $g_{f}$ on the boundary Hilbert space.

%
                                                                                     %

\section{Boundary Quantum Geometry}\label{Implementation of Boundary Quantum Geometry}

We now consider a modification of BF theory.  The modification is obtained by restricting the boundary space $\ch_{\g_v}$ by imposing a certain constraint.  Let us first define this constraint and then discuss the consequences and the motivation of imposing it.

\subsection{Geometric Constraints}\label{Geometric Constraints}

Consider a vertex $v$ and its boundary graph $\g_v$.  For each link $f$, consider the Lie algebra element $\Sigma$ given by
\be
B_f={}^*\Sigma_f+\frac1\gamma \Sigma_f
\label{BS}
\ee
where the star indicates the Hodge dual in the Lie algebra. Consider a node $e$ of the boundary graph  $\g_v$, and let  $f\in\partial e$ be all oriented away from $e$.  Then define
\begin{description}
\item[1. Simplicity Constraint:] There exists a unit vector $(n_e)_I $ for each $e$ such that, for all $f\in\partial e$
\be
(n_e)_I {}^*\Sigma_{f}^{IJ}=0.\label{simple}
\ee

\item[2. Closure Constraint:]
\be
\sum_{f\in\partial e} \Sigma_f^{IJ}=0\label{close}.
\ee
\end{description}
These are the two constraints on which we focus. The main motivation for considering these constraints is the fact that the action of general relativity in the Holst formulation can be written in the form
\be
S_{GR}[e,\omega]=\int B\wedge F[\omega]
\ee
where $\omega$ is an $\slc$ connection,
\be
B={}^*\Sigma+\frac1\gamma \Sigma \label{BSs}
\ee
and
\be
\Sigma^{IJ}=e^I\wedge e^J
\ee
where $e^I$ is the tetrad one form. The restriction $\left.\Sigma_{f}^{IJ}\right|_{\cal B}$ of $\Sigma$ to any space-like boundary $\cal B$ satisfies the conditions:
\be
\left.n_I \Sigma^{IJ}\right|_{\cal B}=0\label{simplec}
\ee
where $n_I$ is the normal to the boundary
and
\be
d \Sigma=0.\label{closec}
\ee
Equations \eqref{BS}, \eqref{simple} and \eqref{close} can be seen as a discrete consequence of equations  \eqref{BSs}, \eqref{simplec} and \eqref{closec}. Here, however, we take the discretized equations  \eqref{BS}, \eqref{simple} and \eqref{close} as our starting point, and study their consequences. A full discussion on the relation of these equations with continuum general relativity will be considered elsewhere.%
\footnote{The Plebanski simplicity constraint implies the constraints given here. However the reverse is not true in general, unless ``shape-matching" conditions \cite{poly} are imposed on each face shared by two polyhedra. We do not demand such shape-matching conditions here. There is some evidences from the large-$j$ behavior of the generalized spinfoam model that non-shape-matching amplitudes are suppressed in the large-$j$ asymptotic \cite{future}.}

The key consequences of these constraints is that they allow $\Sigma$ to determine a classical polyhedral geometry at each node $\t$ of the boundary graph $\g_v$.   (See also \cite{poly}.) This follows from the following
%
%
\begin{Theorem}\label{theo1}
Given an F-valent node $e$ in $\g_v$, let $F$ bivectors $\Sigma_{f}$ satisfy (\ref{simple}) and (\ref{close}). Then there exists a (possibly degenerate) flat convex polyhedron in $\mathbb{R}^3$ with $F$ faces, whose face area bivectors coincide with $\Sigma^{IJ}_{f}$. The resulting polyhedron is unique up to rotation and translation.
\end{Theorem}

\startproof\
Without loss of generality, we fix the unit vector $(n_e)_I=(1,0,0,0)$ (we call this the time-gauge). The simplicity constraint eq.(\ref{simple}) reduces to
\begin{align}
\Sigma_{f}^{0i} = 0.
\end{align}
Hence the surviving components of $\Sigma_{f}^{IJ}$ are $\Sigma^{ij}_{f}$. We denote these nonvanishing components simply by $\Sigma^i_f=\frac12\epsilon^i{}_{jk}B^{jk}$ or $\vec{\Sigma}_f$, in terms of which the closure constraint (\ref{close}) reads
\begin{align}
\sum_{f} \vec{\Sigma}_{f}=0.\label{clob3}
\end{align}

Consider $\vec{\Sigma}_{f}$ as vectors in $\R^3$. Call $|\Sigma_{f}|$ the length of the 3-vector $\vec{\Sigma}_{f}$, and let $\bn:=\bv/\bl$. We first suppose the unit vectors $\bn$ are non-coplanar. Then we recall Minkowski's Theorem \cite{Minkowski}, which states that whenever there are $F$ non-coplanar unit 3-vectors $\bn$ and $F$ positive numbers $A_f$ satisfying the condition
\be
\sum_{f} A_f \vec n_f=0,\label{mink}
\ee
then there exists a convex polyhedron in $\mathbb{R}^3$, whose faces have outward normals $\bn$ and areas $A_f$. And the resulting polyhedron is unique up to rotation and translation.%
\footnote{Imagine the polyhedron immersed in a homogeneous fluid. Eq.\eqref{mink} multiplied by  the pressure  is the sum of the pressure forces acting on the faces, which obviously vanishes.}

 When we apply Minkowski's theorem to our case, we see that the existence of the unit 3-vectors $\bn$ and the lengths $\bl$, as well as the closure constraint eq.(\ref{clob3}), together imply that there is a convex polyhedron in $\mathbb{R}^3$, unique up to translation and rotation, such that each $\bn$ is an outward normal of a face and each $\bl$ is an area of a face. Such a polyhedron can be concretely constructed via LasserreÕs reconstruction algorithm \cite{polyreconstruct}. Let $e^i$ the natural triad in $\R^3$, then the 3-vector $\bv$ can be expressed as an oriented area:
\be
\Sigma^{ij}_{f}=\int_f e^i\wedge e^j.
\ee

Finally, the case of coplanar unit 3-vectors $\bn$ can be obtained as a limit of non-coplanar case, yielding degenerate polyhedra.
\finishproof

This geometrical interpretation equips the variables $e$ and $f$ with a further new meaning: they represent, respectively, polyhedra in a 4d space and faces of these polyedra. See Table 1.

\begin{table}[t]
\begin{center}
\begin{tabular}{|c|c|c|c|}
\hline
{\ } & 2-complex ${\cal K}$ & boundary graph $\gamma_v$& boundary 3d geometry\\
\hline
$e$&  edge& node& polyhedron \\
\hline
$f$& face& link & face of polyhedron\\
\hline
\end{tabular}
\end{center}
\caption{The different geometrical interpretations of the labels $e$ and $f$.}
\end{table}%

The geometrical interpretation in terms of tetrahedra (and now polyhedra) has raised a lively discussion and it is sometimes unpalatable to the more canonical-oriented part of the community. Part of this discussion is based on misunderstanding. The precise claim here is that if we take the diff-invariant Hilbert space of the theory and we \emph{truncate} it to a finite graph (so that the observable algebra is also truncated), then the truncated Hilbert space (with its observables algebra) has a classical limit, and this classical limit can be naturally interpreted as describing a collection of polyhedra. This is well consistent with classical general relativity, because classical general relativity as well admits truncations where the geometry is discretized. Also, this is not inconsistent with the continuous picture for the same reason for which the fact that the truncation of Fock space to an $n$ particle Hilbert space describes discrete particles, is not inconsistent with the fact that Fock space itself describes a (quantized) field.

Let us now see how the constraints translate on the variable $B$ given in \eqref{BS}. We have easily:
\begin{description}
\item[{\bf Simplicity\ Constraint:}]
\be
C_{f}^J=n_I\left(^*B_{f}^{IJ}-\frac{s}{\gamma}B_{f}^{IJ}\right) = 0 \label{pro},
\ee
\item[{\bf Closure\ Constraint:}]
\be
G^{IJ}_e=\sum_{f\in e} B^{IJ}_{f} = 0,\label{clo}
\ee
where $s=+1$ for $Spin(4)$ and $s=-1$ for $\slc$.
\end{description}

Consider a single polyhedron $e$, with the time-gauge $(n_e)_I=(1,0,0,0)$,
and introduce the rotation $L_{f}^j:=\frac{1}{2}\epsilon^{j}{}_{kl}B_{f}^{kl}$ and boost $K_{f}^j:=B_{f}^{0j}$ components of $B^{IJ}_f$. Then  the simplicity constraint (\ref{pro}) becomes simply
\be
\vec K_{f}=s\gamma\ \vec L_{f}\label{key};
\ee
the rotation generators are proportional to the boost generators. The closure constraint (\ref{clo}) can be written as
\begin{subequations}\label{closure}
\begin{align}
\sum_{f\in \partial e} \vec L_{f}&=0,\label{closureL}\\
\mathrm{and}\quad\sum_{f\in \partial e}\vec K_{f}&=0.\label{closureK}
\end{align}
\end{subequations}
where the second, eq.(\ref{closureK}), is redundand, by eq.(\ref{key}).

Let us now move to the quantum theory, and impose the two constraints \eqref{key} and \eqref{closureL} weakly \cite{EPRL,DingYou} on the quantum states. This gives
\begin{description}
\item[{\bf Simplicity\ Constraint:}]
\be
\lag \psi,\vec{K}_{f}\psi'\rag=s\gamma\ \lag \psi,\vec{L}_{f} \psi'\rag\label{GC1}.
\ee
\item[{\bf Closure\ Constraint:}]
\be
\sum_{f\in \partial e}\lag \psi,\vec {L}_{f} \psi'\rag=0\nonumber\\
\sum_{f\in \partial e}\lag \psi,\vec {K}_{f} \psi'\rag=0\label{GC2},
\ee
\end{description}
These equations define a subspace $\ch^E_{\g_v}$ (respectively $\ch^L_{\g_v}$ in Lorentzian case) of the boundary Hilbert space $\ch_{\g_v}$ of BF theory, where the constraints hold weakly. That is, we define $\ch^E_{\g_v}$ as the subspace where these equations hold for any two states $\psi$ and $\psi'$ in a dense domain, for all nodes $s$ of $\g_v$.

\subsection{New Boundary Hilbert Space:\\ Euclidean Theory}\label{New Boundary Hilbert Space: Euclidean Theory}

Let us now construct $\ch^E_{\g_v}$.  Here we first define $\ch^E_{\g_v}$ and then prove that it solves the geometric constraint. We begin with some preliminaries on the structure of the BF boundary Hilbert space. In the Euclidean theory, this space has the following decomposition
\be
\ch_{\gamma_v}=\bigotimes_{f}L^2[Spin(4)]=\bigotimes_{f}\lt[\bigoplus_{\rho_{f}}V_{\rho_{f}}\otimes V_{\rho_{f}}^*\rt].
\label{deco}
\ee
where $V_{\rho}$ denote the representation space for the Spin(4) irrep $\rho$ and $V_{\rho}^*$ is the representation space for the Spin(4) adjoint irrep $\rho^*$.
For each face $f$,  $V_{\rho_{f}}$ and $V_{\rho_{f}}^*$ transforms in
a gauge transformation \eqref{gauge} under the action of $\Lambda_{s_f}$
$\Lambda_{t_f}$, where $s_f$ and $t_f$ are the initial and final points of the link $f$. By regrouping all representations space that transform under the action of the same $\Lambda_{e}$, namely by regrouping the representation spaces associated to the same vertex $e$ of $\g_v$ we can rewrite the decomposition in the form
\begin{equation}
\h_{\gamma_v} = \bigoplus_{\{\rho_{f}\}}\bigotimes_{\t} \bigotimes_{f\in \partial\t}V^{(\t,f)}_{\rho_{f}}\label{vertexdecomp}
\end{equation}
where
\be
V_{\rho_{f}}^{(s_f,f)}&:=&V_{\rho_{f}}\nonumber\\
V_{\rho_{f}}^{(t_ff)}&:=&V_{\rho_{f}}^*
\ee
Therefore the sum over colorings $\rho_{f}$ associates a representation space \begin{equation}
\bigotimes_{f\in \partial\t}V^{(\t,f)}_{\rho_{f}}\label{productrep}
\end{equation}
to each vertex $\t$.
This space can be seen as the quantization of the shapes of a polyhedron with faces having fixed areas, determined by the coloring $\rho_{f}$ \cite{poly}.

Since $Spin(4)\sim SU(2)_+\times SU(2)_-$, a unitary irrep of $Spin(4)$ is given by a tensor product of two SU(2) irreps. $V_\rho=V_{j^+}\otimes V_{j^-}$ with spins $j^+$ and $j^-$. We can characterize $\rho$ by $\rho=(p,k)$, where
\be
 p=j^++j^-, \hspace{3em}  k=j^+-j^-.
\ee
The $SU(2)_\pm$ subgroups of $Spin(4)$ are its canonical self-dual and antiself dual components, generated by $\vec L\pm\vec K$, and should not be confused with the (non-canonical) SU(2) subgroup generated by $\vec L$, used to pick a time gauge. If we decompose $V_\rho=V_{p,k}$ in irreducibles of SU(2), we have
\begin{equation}
V_{p,k}=V_{j^+} \otimes V_{j^-} = \bigoplus_{j=|k|}^{p}V_j^{p,k}.\label{CG}
\end{equation}

We now define $\ch^E_{\g_v}$.  In the representation space $V_{p,k}$, pick the $V_j^{p,k}$ subspace (in the decomposition above), where $j$ is defined by
\be
p&=&j+r \label{p}\\
k&=&\g j-r
\ee
By doing so, we obtain the subspace $V_j^{\gamma j +r,j-r}$ in each $V_{p,k}$.
By restricting in this manner all the $V_{\rho_f}$ subspaces in \eqref{deco} we obtain a subspace of $\ch_{\gamma_v}$. We define the non-gauge-invariant new boundary space to be this subspace.
That is
\be
 \bigoplus_{\{j_{f},r_f\}}\bigotimes_{\t} \bigotimes_{f\in \t}(V^{ j_f +r_f,\gamma j_f-r_f}_{j_f})^{(\t,f)}\label{vertexdecomp}
\ee
where the sum is over non-negative half-integers $j_f$ and $r_f$.
The possible coloring in $\ch^E_{\g_v}$ are labelled by the two non-negative half-integer quantum numbers $j_f$ and $r_f$. The quantum number $j_f$ characterizes the SU(2) spin of the representation and is easily identified with the corresponding  LQG quantum number which is associated to each link of the graph. $r_f$ is a new quantum number, also associated to each link of the graph.

Notice also that \eqref{p} restricts also the possible values of $j$ and $r$ to those for which  $p=\gamma j +r$ is half integer. This awkward feature of the Euclidean case disappears in the Lorentzian theory.

We can translate all this in terms of the $(j^+,j^-)$ notation. This gives
\be
j^+=\f{1+\g}{2}j\ \ \ \ \text{and}\ \ \ \ j^-=\f{1-\g}{2}j+r\label{ksimple}
\ee
and the  modified $\g$-simplicity relation %
\footnote{The $Spin(4)$ irreps for a given Barbero-Immirzi parameter $\g$, should be such that
\be
r=\frac{(1+\g)j^--(1-\g)j^+}{1+\g}
\ee
is a non-negative integer, and satisfy
\be
0\leqslant r\leqslant j^++j^--|j^+-j^-|
\ee
implying
\be
\frac{|1-\g|}{1+\g}j^+\leqslant j^-\leqslant j^+ \ \ \ \ \text{or}\ \ \ \ j^+\leqslant j^-
\leqslant\frac{3+\g}{1+\g}j^+.
\ee
}
%
\be
(1-\g)j^+=(1+\g)(j^--r).
\ee



Next, we define the gauge invariant new boundary space. Consider the diagonal actions of $h\in SU(2)$ on each product representation space eq.(\ref{productrep}) at each $\t$. We denote the invariant subspaces under this actions by
\be
\Fraki_\t^{\{j_{f}\}}=\text{Inv}_{SU(2)}\lt[\bigotimes_{f\in \t}(V_{j_{f}}^{ j_f+r_f,\gamma j_f-r_f})^{(\t,f)}\rt]
\ee
The  gauge invariant new boundary Hilbert space is defined by
\begin{equation}
\ch^E_{\g_v}:=\bigoplus_{\{j_{f},r_f\}}\bigotimes_{\t} \Fraki_\t^{\{j_{f}\}}.
\end{equation}

An orthonormal basis in $\ch^E_{\g_v}$ can be constructed as follows.  Given a polyhedron $\t$ with $F$ faces, we assign at $\t$ an $F$-valent $SU(2)$ intertwiner $i^{A_1\cdots A_F}_\t$ associated with $F$ $SU(2)$ irreps $j_{f}$, $f=1,\cdots F$. An orthonormal basis is then defined by the following functions on $[Spin(4)]^{|E(\g_v)|}$
\be
&&
\hspace{-3mm}
T^E_{\g_v,j_f,r_{f},i_\t}(g_{f})\!=\!\prod_f\sqrt{[(1\!+\!\g)j_{f}+1]
[({1\!-\!\g})j_{f}+2r_{f}+1]}\nonumber\\
\hspace{-3mm}
&&
\hspace{-3mm}
\prod_{\t}\bigg[i^{A_{\t1}\cdots A_{\t F}}_\t
\prod_{f\in e}
C_{A_{\t1}}^{m^+_{\t f}m^-_{\t f}}\bigg]
\prod_f \bigg[\eps^{n^+_{\t f}n^+_{\t' f}} \eps^{n^-_{\t f}n^-_{\t' f}}\bigg]\label{TE}
\nonumber\\
&&
\hspace{-3mm}
\prod_{(e,f)}\bigg[D^{\f{1+\g}{2}j_{f}}_{m^+_{\t f}n^+_{\t f}}(g^+_{\t f})D^{\f{1-\g}{2}j_{f}+r_{f}}_{m^-_{\t f}n^-_{\t f}}(g^-_{\t f}) \bigg]
\ee
here $g_{\t f}=(g_{\t f}^+,g_{\t f}^-)\in Spin(4)$, $D^j(g)$ is the representation matrix of the $SU(2)$ irrep $j$, and $C_{A_{\t f}}^{m^+_{\t f}m^-_{\t f}}$ denotes the Clebsch-Gordan coefficient ($A_f=-k_f,\cdots,k_f$)
\be
&&\lag \f{1+\g}{2}j_{f},\ \f{1-\g}{2}j_{f}+r_{f};\ j_f,\ A_{\t f}\ \Big|\right. \\
&&\left. \hspace{5em} \Big|
\ \f{1+\g}{2}j_{f},\ m^+_{\t f};\ \f{1-\g}{2}j_{f}+r_{\t f},\ m^-_{\t f}\rag.\nonumber
\ee
$\eps^{n^\pm_{\t f}n^\pm_{\t' f}}$ are the unique 2-valent $SU(2)$ intertwiners with representations $j^+_f=\f{1+\g}{2}j_{f}$ and $j^-=\f{1-\g}{2}j_{f}+r_{f}$ respectively. Thus $T^E_{(\g_v,j_f,r_{f},i_\t)}$ is essentially a function over $g_{f}=g_{\t f}g_{f\t'}$. Note that if we ask the quantum numbers $r_f$ to be some fixed integers, then the spin-network functions $T^E_{(\g_v,j_f,r_{f},i_\t)}$ can be equivalently considered as an $SU(2)$ spin-network functions, thus the boundary Hilbert space is spanned by $SU(2)$ spin-networks, as the case of LQG kinematical Hilbert space.

We are now ready to prove our first main result.\\

\begin{Theorem}
The Hilbert space $\ch^E_{\g_v}$ solves the geometric constraint (\ref{GC1}-\ref{GC2}), with $s=1$.
\end{Theorem}

\startproof  The closure constraint \Ref{GC2} follows immediately since the states in $\ch^E_{\g_v}$ is invariant under the diagonal $SU(2\partial $ gauge transformation $(g^+_{\t f},g^-_{\t f})\mapsto(h_\t g^+_{\t f},h_\t g^-_{\t f})$ at each $\t$ (the constraint is even solved strongly). The nontrivial proof is for the simplicity constraint \Ref{GC1}. Define the self-dual/anti-self-dual operators:
\begin{align}
\vec {J}_{f}^{\pm}:=&\f{1}{2}(\vec {L}_{f}\pm
\vec {K}_{f})
\end{align}
then (\ref{GC1}) reads
\begin{align}
(1-\gamma)\lag \psi,\vec{J}_{f}^{+}\psi'\rag_E-(1+\gamma)\lag f,\vec{J}_{\psi}^{-}\psi'\rag=0\label{propm}.
\end{align}
The operators $\vec{J}_{f}^{\pm}$ on $L^2(Spin(4))$ act on individual $V^{(\t,f)}_{\rho_f}$ (see, e.g. Sec.32.2 of \cite{thiemannbook}). Therefore we only need to show that in each Clebsch-Gordan subspace $V_j^{\rho=(j^+,j^-)}$, with $j^+\equiv\f{1+\g}{2}$ and $j^-\equiv\f{1-\g}{2}k+r$, the following relation holds for all pairs $\Phi,\Psi$ of vectors
\be
(1-\gamma)\bra{\Psi}\vec{{J}}^{+}\ket{\Phi}-(1+\gamma)\bra{\Psi}\vec{{J}}^{-}\ket{\Phi}=0\label{mepm}
\ee
where $\lag\ |\ \rag$ is the Hermitian inner product on the $Spin(4)$ irrep $V_{\rho=(j^+,j^-)}$.

To evaluate these matrix elements, we use the explicit representation of the vectors as multi-spinors. The vectors in the $SU(2)$ irrep $V_j$ can be represented as totally symmetric spinorial tensors with $2j$ spinor indices. The generators of $SU(2)$ are then Pauli matrices $\sigma^A_i{}_B$  acting on each index, followed by a sum. A state $\ket{\Phi}$ in $\h_{j}$ in the Clebsch-Gordan subspace $V_j^{j^+,j^-}\subset V_{j^+}\otimes V_{j^-}$ can be expressed by ($A_i,B_i=1,2$)
\be
\Phi^{A_1...A_{2j^+},B_1...B_{2j^-}}&=& \label{phi}\\
&&\hspace{-5em}\epsilon^{A_1B_1}...\epsilon^{A_{r}B_{r}}\phi^{A_{r+1}...A_{2j^+},B_{r+1}...B_{2j^-}},\nonumber
\ee
with complete symmetrization of all $(A_1,...A_{2j^+})$ indices understood and the same for the $(B_1,...,B_{2j^-})$ indices.
The action of $\vec J^{-}$ on the state $\Phi$ in \Ref{phi}, can then be computed explictly, giving
\begin{widetext}
\be
&&J^{-i}\Phi^{(A_1...A_{2j^+})(B_1...B_{2j^-})}
=\sum_{p=1}^{2j^-}\sigma_i^{B_p}{}_{\widetilde{B}_p}\Phi^{(A_1...A_{2j^+})(B_1...\widetilde{B}_p...B_{2j^-})}\\
&&=\sum_{p=1}^{r}\sigma_i^{B_p}{}_{\widetilde{B}_p}\epsilon^{A_1B_1}...\epsilon^{A_p\widetilde{B}_p}...\epsilon^{A_{r}B_{r}}\phi^{(A_{r+1}...A_{2j^+}B_{r+1}...B_{2j^-})}+\sum_{p=r+1}^{2j^-}\sigma_i^{B_p}{}_{\widetilde{B}_p}\epsilon^{A_1B_1}...\epsilon^{A_{r}B_{r}}\phi^{(A_{r+1}...A_{2j^+}B_{r+1}...\widetilde{B}_{p}...B_{2j^-})}\nonumber\\
&&=-\sum_{p=1}^{r}\sigma_i^{A_p}{}_{\widetilde{A}_p}\epsilon^{A_1B_1}...\epsilon^{\widetilde{A}_p{B}_p}...\epsilon^{A_{r}B_{r}}\phi^{(A_{r+1}...A_{2j^+}B_{r+1}...B_{2j^-})}+\sum_{p=r+1}^{2j^-}\sigma_i^{B_p}{}_{\widetilde{B}_p}\epsilon^{A_1B_1}...\epsilon^{A_{r}B_{r}}\phi^{(A_{r+1}...A_{2j^+}B_{r+1}...\widetilde{B}_{p}...B_{2j^-})}\nonumber
\ee
where in the third step, we use the identity $\sig_i^B{}_{\widetilde{B}}\epsilon^{A\widetilde{B}}=-\sigma_i^{A}{}_{\widetilde{A}}\epsilon^{\widetilde{A}{B}}$ coming from the $\slc$ invariance of $\eps^{AB}$. Then the matrix elements of $\vec J^{- }$ are
\be
\bra{\Psi}J^{-i}\ket{\Phi}
&=&-\sum_{p=1}^{r}\sigma_i^{A_p}{}_{\widetilde{A}_p}\Psi_{(A_1...A_p...A_{2j^+})(B_1...B_{2j_-})}\Phi^{(A_1...\widetilde{
A}_p...A_{2j^+})(B_1...B_{2j_-})}\nonumber\\
&&+\sum_{p=r+1}^{2j^-}\sigma_i^{B_p}{}_{\widetilde{B}_p}
\epsilon_{A_1B_1}...\epsilon_{A_{r}B_{r}}\psi_{(A_{r+1}...A_{2j^+}B_{r+1}...{B}_{p}...B_{2j^-})}
\epsilon^{A_1B_1}...\epsilon^{A_{r}B_{r}}\phi^{(A_{r+1}...A_{2j^+}B_{r+1}...\widetilde{B}_{p}...B_{2j^-})}\nonumber\\
&=&(-r)\sigma_i^{A_{2j^+}}{}_{\widetilde{A}_{2j^+}}\Psi_{(A_1...A_{2j^+})(B_1...B_{2j_-})}\Phi^{({A}_1...\widetilde{A}_{2j^+})(B_1...B_{2j_-})}\nonumber\\
&&+(2j^--r)\sigma_i^{A_{2j^+}}{}_{\widetilde{A}_{2j^+}}
\epsilon_{A_1B_1}...\epsilon_{A_{r}B_{r}}\psi_{(A_{r+1}...A_{2j^+}B_{r+1}...B_{2j^-})}
\epsilon^{A_1B_1}...\epsilon^{A_{r}B_{r}}\phi^{(A_{r+1}...\widetilde{A}_{2j^+}B_{r+1}...B_{2j^-})}\nonumber\\
&=&2(j^--r)\sigma_i^{A_{2j^+}}{}_{\widetilde{A}_{2j^+}}\Psi_{(A_1...A_{2j^+})(B_1...B_{2j_-})}\Phi^{({A}_1...\widetilde{A}_{2j^+})(B_1...B_{2j_-})}
\ee
Similarly,
\begin{align}
\bra{\Psi}J^{+i}\ket{\Phi}=2j^+\sigma_i^{A_{2j^+}}{}_{\widetilde{A}_{2j^+}}\Psi_{(A_1...A_{2j^+})(B_1...B_{2j_-})}\Phi^{({A}_1...\widetilde{A}_{2j^+})(B_1...B_{2j_-})}.
\end{align}
Then eq.\Ref{mepm} follows immediately
\be
&&(1-\gamma)\bra{\Psi}{{J}}^{(+)i}\ket{\Phi}-(1+\gamma)\bra{\Psi}{{J}}^{(-)i}\ket{\Phi}\nonumber\\
&&\quad \quad = 2\lt[(1-\g)j^+-(1+\g)(j^--r)\rt]\sigma_i^{A_{2j^+}}{}_{\widetilde{A}_{2j^+}}\Psi_{(A_1...A_{2j^+})(B_1...B_{2j_-})}\Phi^{({A}_1...\widetilde{A}_{2j^+})(B_1...B_{2j_-})}\nonumber\\
&&\quad \quad = 0\label{mepm0}
\ee
which proves the simplicity constraint eq.\Ref{GC1}.
\finishproof
\end{widetext}

\subsection{New Boundary Hilbert Space: \\ Lorentzian Theory}\label{New Boundary Hilbert Space: Lorentzian Theory}

Now we turn to the case of $G=\slc$. In this case the decomposition of the Hilbert space reads
\be
\ch_{\g_v}&=&\bigotimes_{f}L^2\Big(\slc,\rmd\mu_H\Big)\label{HSL2C0}\\\nonumber
&=&\bigotimes_{f}\bigoplus_{k_f=\mathbb{N}/2}\int_{\mathbb{R}}^\oplus\rmd p_f\lt(p^2_f+k^2_f\rt)\ V_{(k_f,p_f)}\otimes V_{(k_f,p_f)}^*
\ee
where $k_f$ are still non-negative half-integers but $p_f\in\mathbb{R}$ is now a real number. Here $\int^\oplus$ denotes a direct integral decomposition \cite{gelfand} (see also Chapter 30 of \cite{thiemannbook}). $V_{(k,p)}$ denotes the unitary irrep of $\slc$ in the principal series, and $V_{(k,p)}^*$ denotes the adjoint irrep. We can then proceede as in the EUclidean theory.
The BF boundary Hilbert space reads
\be
\ch_{\g_v}=\bigoplus_{\{k_f\}}\prod_{f}\int^\oplus_{\mathbb{R}}\rmd p_f\prod_f\lt(p^2_f+k^2_f\rt)\bigotimes_{\t}\bigotimes_{f\in \t}V_{(k_{f},p_f)}^{(\t,f)}\label{HSL2C}
\ee
The representation space $V_{(k,p)}$ is infinite-dimensional and can be decomposed into $SU(2)$ irreps (irreps of the subgroup generated by $\vec {L}$), i.e.
\be
V_{(k,p)}=\bigoplus_{j=k}^\infty V_j^{k,p}.
\ee
This time we introduce the two parameters $j$ and $r$ by
\be
p&=&\g j \ \frac{j+1}{j-r}\label{rsimple},\\
k&=&j-r.
\ee
and we define the new boundary space by restricting each $V_{(k,p)}$ to its
$ V_j^{k,p}$ subspace satisfying \eqref{rsimple}.
This time $p$ does not need to be half-integer, therefore \eqref{rsimple} can be solved for any $j$. The new quantum numbers associated to each face are $j_f$ and $r_f$, each being a nonnegative half integer.

As before, we consider the diagonal $SU(2)$ action at each $\t$
for all $h_\t\in SU(2)$. The invariant subspace under this action is
\be
\Fraki_\t^{j_{f}}=\mathrm{Inv}_{SU(2)}\lt[\bigotimes_{f\in \t}\left(V^{\frac{\g j_f(j_f+1)}{j_f-r_f},j_f-r_f}\right)^{(\t,f)}
\rt]
\ee
The new boundary Hilbert space is defined by a product of these invariant subspaces over all the polyhedra $\t$, followed by a sum over all the possible $j_f$ and $r_f$:
\be
\ch^L_{\g_v}:=\bigoplus_{\{r_f,j_f\}}\ \bigotimes_{\t}\Fraki_\t^{j_{f}}
\ee
where $j_f$ and $k_f$ are non-negative half-integers with constraints (1) $j_f\geq r_f$. $\ch^L_{\g_v}$ is a direct sum over a set of subspaces contained in the fiber Hilbert spaces of $\ch_{\g_v}$ (see eq.(\ref{HSL2C0})), thus has well-defined inner product.

An orthonormal basis is constructed as follows. Consider the oriented boundary graph $\g_v$.
%
Given a $F$-valent vertex/polyhedron $\t$, we assign it an intertwiner $i_\t^{A_1\cdots A_F}$ associated with $F$ spins $j_f$, $f=1,\cdots,F$
\be
i_\t\in\mathrm{Inv}\lt[\bigotimes_{\overrightarrow{(\t,f)}\ \text{outgoing}}V_{j_f}\bigotimes_{\overrightarrow{(\t,f)}\ \text{incoming}}V_{j_f}^*\rt]
\ee
An orthogonal basis in $\ch^L_{\g_v}$ is given by the following functions (distributions) on $\slc$
\be
&&T^L_{(\g_v,j_f,r_f,i_\t)}(g_{f})=\\ \nonumber
&&\quad \prod_{\t}i_\t^{A_{\t1}\cdots A_{\t F}}\prod_{(\t,\t')}\Pi^{(\frac{\g j_f(j_f+1)}{j_f-r_f},j_f-r_f)}_{j_fA_{\t f},j_fA_{\t' f}}(g_{f})\label{TL}
\ee
here $\Pi^{(p,k)}$ denotes the representation matrix in $\slc$ irrep labeled by $(p,k)$. All the $A_{\t f}$ indices of the representation matrices are contracted with the $A_{\t f}$ indices of the intertwiners.

The new boundary Hilbert space $\ch^L_{\g_v}$ is \emph{not} a subspace of the BF boundary Hilbert space $\ch_{\g_v}$, because $T^L_{(\g_v,j_f,r_f,i_\t)}$ are constructed by $\Pi^{(k,p)}$ which are distributions. In order to check the geometric constraints Eqs.(\ref{GC1}) and (\ref{GC2}) on $\ch^L_{\g_v}$, we have to compute the (dual) action of the bivector operator on the distributions $T^L_{(\g_v,j_f,k_f,i_\t)}$. Fortunately the Hilbert space $L^2\big(\slc)$ has the structure of direct integral decomposition (see eq.(\ref{HSL2C0})). Then the (dual) action of the bivector operators $\vec\hat{K}$ and $\vec\hat{L}$ gives the actions of Lie algebra generators $\vec L$ and $\vec K$ on each fiber Hilbert space $V_{(k,p)}$.

We are now ready to proove our second main result\\

\begin{Theorem}
The Hilbert space $\ch^L_{\g_v}$ solves the geometric constraint (\ref{GC1},\ref{GC2}), with $s=-1$.
\end{Theorem}

\startproof Closure constraint follows immediately and strongly by the diagonal $SU(2)$ invariance at each polyhedron $\t$.  We only need to consider a single irrep $V_{(k,p)}$ ($p=\frac{\g j(j+1)}{k}$) because $\vec L$ and $\vec K$ leave it invariant and, different $(p,k)$'s label orthogonal subspaces in $\ch^L_{\g_v}$.

A canonical basis in $V_{(p,k)}$ is obtained diagonalizing the Casimir operators $J\cdot J, {}^*J\cdot J, L\cdot L$ and $L^3$. The basis can be denoted $\ket{(p,k);j,m}$ or simply as $\ket{j,m}$ since we only consider a single irrep. On this canonical basis, the generators act in the
following way \cite{gms}:
\be
L^3\ket{j,m}&=& m\ket{j,m}, \nonumber \\
L^+\ket{j,m}&=& \sqrt{(j+m+1)(j-m)}\ket{j,m+1}, \nonumber\\
L^-\ket{j,m}&=& \sqrt{(j+m)(j-m+1)}\ket{j,m-1}, \nonumber \\
K^3\ket{j,m}&=& -\alpha_{(j)}\sqrt{j^2-m^2}\ket{j-1,m}-\beta_{(j)}m\ket{j,m}\nonumber\\
&& +\alpha_{(j+1)}\sqrt{(j+1)^2-m^2}\ket{j+1,m}, \nonumber \\
K^+\ket{j,m}&=& -
\alpha_{(j)}\sqrt{(j-m)(j-m-1)}\ket{j-1,m+1}\nonumber\\
&&-\beta_{(j)}\sqrt{(j-m)(j+m+1)}
\ket{j,m+1} \label{lorentzrep}\nonumber \\
&&-\alpha_{(j+1)}\sqrt{(j+m+1)(j+m+2)}\ket{j+1,m+1},\nonumber \\
K^-\ket{j,m}&=&
\alpha_{(j)}\sqrt{(j+m)(j+m-1)}\ket{j-1,m-1}\nonumber\\
&&-\beta_{(j)}\sqrt{(j+m)(j-m+1)}
\ket{j,m-1} \nonumber \\
&&+\alpha_{(j+1)}\sqrt{(j-m+1)(j-m+2)}\ket{j+1,m-1}, \nonumber
\ee
where
\be
 L^{\pm}=L^1\pm iL^2,\qquad K^{\pm}=K^1\pm iK^2
 \ee
 and
\be&\alpha_{(j)}=\frac{i}{j}\sqrt{\frac{(j^2-k^2)(j^2+p^2)}{4j^2-1}}, \qquad\beta_{(j)}=\frac{kp}{j(j+1)}
\ee
Using these equations, one can check directly that
\begin{align}
\bra{j,m'}\big(K^i+\beta_{(j)} L^i\big)\ket{j,m}=0.\label{beta}
\end{align}
which is nothing but
\begin{align}
\bra{j,m'}\big(K^i+ \gamma L^i\big)\ket{j,m}=0.
\end{align}
because $pk={\g j(j+1)}$.

\finishproof

\subsection{Quantum Polyhedral Geometry}\label{Quantum Polyhedral Geometry}

In this section we show that the boundary Hilbert space $\ch^E_{\g_v}$ and $\ch^L_{\g_v}$ carries a representation of quantum polyhedral geometry, consistent with the classical polyhedral geometry that we have discussed in Section \ref{Geometric Constraints}. Recall that we defined two different bivectors $J_{\t f}^{IJ}$ and $\Sigma_{\t f}^{IJ}$ related by
\be
B^{IJ}_{f}=\lt({}^*\Sigma_f+\frac{1}{\g}\Sigma_f\rt)^{IJ}_{\t f}
\ee
Theorem \ref{theo1} states that classically, the geometric constraint of $B^{IJ}_{f}$ implies that $B^{IJ}_{f}$ is the area bivector of a face $f$ of a polyhedron $e$. On the BF boundary Hilbert space $\ch_{\g_v}$ the bivector $B^{IJ}_{f}$ is quantized to be the left invariant vector field ${J}^{IJ}_{f}$. Inverting the above equation, we can write the quantum operator corresponding to $\Sigma$ (which we indicate with the same symbol) as
\be
\Sigma^{IJ}_{f}:=\frac{\g^2}{\g^2-s}\lt({}^*{J}_{\t f}^{IJ}-\frac{1}{\gamma}\ {J}_{\t f}^{IJ}\rt)
\ee
Give a polyhedron/vertex $\t$ of the boundary, if we choose the unit vector $(n_\t)_I=(1,0,0,0)$, then the simplicity constraint implies the vanishing of $\Sigma^{0j}_{f}$ for each face $f$. That is, the matrix elements of the operators ${\Sigma}^{0i}_{f}$ vanish on $\ch^E_{\g_v}$ and $\ch^L_{\g_v}$, thus we consider them as vanishing operators on $\ch^E_{\g_v}$ or $\ch^L_{\g_v}$. The nontrivial operator on $\ch^E_{\g_v}$ and $\ch^L_{\g_v}$ is
\be
{\Sigma}^{i}_{f}\equiv\frac12 \epsilon^i{}_{jk}{\Sigma}^{jk}_{f}=\frac{\g^2}{\g^2-s}\lt(\hat{K}_{f}^{i}-\frac{1}{\gamma}\hat{L}_{f}^i\rt)
\ee
Because of the quantum simplicity constraint (\ref{GC1}), we can
 identify $\hat{K}_{\t f}^{i}$ with $s\g\vec{L}_{\t f}$ on the dense domain of the new boundary Hilbert space, as far as the matrix elements of the operators are concerned. Thus, in the sense of their matrix element
\be
\vec {\Sigma}_{f}=s\g \ \vec {L}_{f}
\ee
By the $SU(2)$ gauge invariance, then
\be
\sum_{f\in\partial\t}\hat{\Sigma}_{f}=0
\ee
(with all $f$'s oriented out of $e$.) Consider now a family of coherent states that makes the spread of these operators small.  These coherent states are then characterized by eigenvalues of $\vec \Sigma_f$ that satisfy the equation above. By Minkowski theorem, they determine a polyhedron $e$ at each vertex.  $\vec {\Sigma}_{\t f}$ represents the normal to face area of the polyhedron $\t$, normalized so that its norm is the area of the face \cite{poly}.
The area operator for a face $f$ (in units that $8\pi\ell_p^2=1$ \cite{simple}) is then
\be
\hat{A}_f=\g\sqrt{\hat{L}_{\t f}^i\hat{L}_{\t f}^i}=\g \sqrt{j_f(j_f+1)}.
\ee
It is clear that the area operator doesn't depend on the orientation of the face. Thus the two areas of the two faces of the two polyhedra $e$ and $e'$ that are determined by the same face $f$ are equal. (Recall that the one of the two is determined by the left invariant vector field $J$ and the other by the right invariant vector field $R$, since $R_f=J_{f^{-1}}$.)

At fixed values of the areas, the shapes of the polyhedra is described by the intertwiner spaces at each $e$.  We recall that an over-complete basis in these spaces is formed by the Livine-Speziale coherent intertwiners \cite{FK}
\be
||\vec{j},\vec{n}\rangle:=\int_{SU(2)}\rmd\mu_H(g)\prod_{f\subset\t}D^{j_f}(g)|j_f,n_f\rangle
\ee
These can be labeled  \cite{FC}  by the elements in $\times_fS^2/\slc$. Thinking of $S^2$ as the compactified complex plane of $z_f$, a coherent intertwiner is determined by $F$ quantum area $j_f$ and $F-3$ complex cross-ratios $\vec{Z}$
\be
Z_k=\frac{(z_{k+3}-z_1)(z_2-z_3)}{(z_{k+3}-z_3)(z_2-z_1)}
\ee
which are invariants of $\slc$. The space of these cross-ratio $\times_fS^2/\slc$ can be identified \cite{Freidel:2009nu} with the Kapovich and Millson phase space $\cs_F$ \cite{KM}, which is also \emph{the space of shapes of polyhedra at fixed areas $j_f$}. Thus, we can label the coherent intertwiner by $||\vec{j},\vec{Z}\rangle$, in variables that relate directly to the shape of the polyhedron. The resolution of identity in the intertwiner space can be expressed as a integral over the Kapovich and Millson phase space $\cs_F$, i.e.
\be
\mathbf{1}_{\ci(\vec{j})}=\int_{\cs_F}\rmd\mu(\vec{Z})\ ||\vec{j},\vec{Z}\rangle\ \langle\vec{j},\vec{Z}||
\ee
where the explicit expression of the measure $\rmd\mu(\vec{Z})$ is given in \cite{FC}. Finally the volume operator for a polyhedron can be defined as in \cite{poly}, in terms of the classical volume of a polyhedron and the coherent intertwiner.

Notice that the quantum polyhedral geometry doesn't depend on the quantum numbers $r_f$.  The quantum numbers $r_f$ don't affect the quantum 3-geometry on the boundary.

\section{Amplitudes}\label{New spinfoam Model}

\subsection{Vertex Amplitude: Euclidean theory}

If we take BF theory and restrict all vertex-boundary spaces to $\ch^E_{\g_v}$ (or $\ch^L_{\g_v}$) we obtain a new dynamical model.  Here we give explicitly its vertex and face amplitude.  Let's start with the Euclidean case. The BF vertex amplitude can be written in the holonomy representation: (each edge joining at $v$ is uniquely determined by a vertex/polyhedron $\t$ on the boundary) reads
\be
A_v(g_{f})&=&\!\!\!\sum_{j^\pm_f,i^\pm_\t}
\prod_f\sqrt{2j_f^++1}\sqrt{2j_f^-+1}\nonumber\\
&& A_v(j^+_f,j^-_f;i^+_\t,i^-_\t)
T^{BF}_{\g_v,j^\pm_f,i^\pm_\t}(g_{f})\label{Avg}
\ee
Here
\be
A_v(j^+_f,j^-_f;i^+_\t,i^-_\t)=\tr\left(\bigotimes_{e\in v}I^\dagger_{e}\right)
\ee
where $ I=(i^+,i^-)$ and we assume the valence of $v$ is $n$. $T^{BF}_{(\g_v,j^\pm_f,i^\pm_\t)}\in\ch_{\g_v}$ is a $Spin(4)$ spin-network function on the boundary graph $\g_v$
\be
T^{BF}_{\g_v,j^\pm_f,i^\pm_\t}(g_{f}):=
T_{\g_v,j^+_f,i^+_\t}(g^+_{f})
T_{\g_v,j^-_f,i^-_\t}(g^-_{f})
\ee
where
\be
T_{\g_v,j_f,i_\t}(g_{f})&=&\prod_f\sqrt{2j_f+1}\prod_{\t}\bigg[(i_\t)^{\{m_{\t f}\}}\bigg]
\\ \nonumber
&&\ \ \prod_{(\t,f)}\bigg[D^{j_f}_{m_{\t f}n_{\t f}}(g_{\t f})\ \bigg]\bigotimes_f \bigg[\eps^{n_{\t f}n_{\t' f}}\ \bigg]
\ee
The vertex amplitude eq.(\ref{Avg}) is a distribution of the boundary Hilbert space $\ch_{\g_v}$, i.e. there is a dense domain of $\ch_{\g_v}$ spanned by the spin-network functions $T^{BF}_{(\g_v,j^\pm_f,i^\pm_\t)}$, such that $A_v(g_{\t\t'})$ lives in the algebraic dual of this dense domain. After imposing the geometric constraint, we restrict ourself to the subspace $\ch^E_{\g_v}$. Such a restriction results in a (dual) projection of the vertex amplitude $A_v$, i.e. we obtain
\be
A_v^E(g_{f})=\!\!\sum_{j_f,r_f,i_\t}\!\!\lag T_{\g_v,j_f,r_f,i_\t},A_v\rag\, T^E_{\g_v,j_f,r_f,i_\t}(g_{f})
\ee
where $T_{\g_v,j_f,r_f,i_\t}$ is a orthonormal basis of $\ch^E_{\g_v}$ (recall eq.(\ref{TE})), and $\lag\ ,\ \rag$ is the inner product of the BF boundary Hilbert space $\ch_{\g_v}$. The evaluation of $A_v^E$ is straightforward:
\be
&&A_v^E(g_{f})=\sum_{j_f,r_f,i_\t}\prod_f\sqrt{2j^+_{f}+1}\sqrt{2j^-_{f}+1}
\\ \nonumber
&&\quad  \sum_{i^+_\t,i^-_\t}A_v\lt(j^+_{f},\ j^-_{f};\ i^+_\t,\ i^-_\t\rt)\prod_{\t}f_{i_\t^+,i_\t^-}^{i_\t}\ T^E_{\g_v,j_f,r_f,i_\t}(g_{\t\t'})
\ee
where we write $j^+\equiv\frac{1+\gamma}{2}j$ and $j^-\equiv\frac{1-\gamma}{2}j+r$ and for each $F$-valent boundary polyhedron/vertex
\be
f_{i_\t^+,i_\t^-}^{i_\t}=\overline{i^{A_{\t1}\cdots A_{\t F}}_\t} C_{A_{\t1}}^{m^+_{\t1}m^-_{\t1}}\cdots C_{A_{\t F}}^{m^+_{\t F}m^-_{\t F}}\nonumber\\
(i^{+}_\t)_{m^+_{\t 1}\cdots m^+_{\t F}}(i^{-}_\t)_{m^-_{\t 1}\cdots m^-_{\t F}}
\ee
Then in the $(j_f,r_f,i_\t)$-spin-network representation, the vertex amplitude is
\be
A^E_v(j_f,r_f,i_\t)=\!\sum_{i^+_\t,i^-_\t}\!\!A_v\!\lt(j^+_{f},j^-_f;\ i^+_\t,\ i^-_\t\rt)\prod_{\t}f_{i_\t^+,i_\t^-}^{i_\t}
\ee
which nontrivially depends on the quantum numbers $r_f$ via the definition of $j^-_f$.

There is another way to write this vertex amplitude in $(j_f,r_f,i_\t)$-spin-network representation. Define a map $I^{\{r_f\}}_E$ from $SU(2)$ intertwiners to $Spin(4)$ intertwiners, depending on the quantum numbers $r_f$. Given an $F$-valent $SU(2)$ intertwiner $i_\t$ with spins $k_1,\cdots,k_F$, let
\be
&& I^{r_f}_E:\ i_\t\mapsto I^{r_f}_E(i_\t)=i^{A_{\t1}\cdots A_{\t F}}_\t C_{A_{\t1}}^{n^+_{\t1}n^-_{\t1}}\cdots C_{A_{\t F}}^{n^+_{\t F}n^-_{\t F}}\nonumber\\
&&\int\rmd g^+\rmd g^-\ \prod_{f\in e} D^{\f{1+\g}{2}j_{k}}_{m^+_{\t f}n^+_{\t f}}(g^+)\ D^{\f{1-\g}{2}j_{f}+r_{f}}_{m^-_{\t f}n^-_{\t f}}(g^-)
\ee
Given an edge $e\in E(\ck)$, we associate an intertwiner $I^{\{r_f\}}_E(i_\t)$ to the inital point of the edge $e$, and a dual intertwiner $I^{\{r_f\}}_E(i_\t)^\dagger$ to the final point of $e$. Then the vertex amplitude $A^E_v$ can be written a spinfoam trace of the intertwiners $I^{\{r_f\}}_E(i_\t)$
\be
A^E_v(k_f,r_f,i_\t)=\tr\lt(\bigotimes_{e\in v}I_E^{\{r_f\}}(i_{\t})^\dagger\rt)
\ee
where we have again assumed that all the edges joining at $v$ are oriented towards $v$.

\subsection{Vertex Amplitude: Lorentzian theory}

The Lorentzian vertex amplitude can be defined in the same manner. The $\slc$ BF vertex amplitude is expressed in the holonomy representation as a distribution
\be
A_v(g_{f})&=&\sum_{k_f,I_\t}\int\prod_f\rmd p_f\prod_f\lt(k_f^2+p_f^2\rt)\\ \nonumber
&&\quad  A_v\Big(p_f,k_f;I_\t\Big)\ T^{BF}_{\g_v,(k,p)_f,(\mathbf{l},\mathbf{n})_\t}(g_{f})
\ee
where
\be
A_v\Big(p_f,k_f;I_\t\Big)=\tr\lt(\bigotimes_e I_e^\dagger\rt)
\ee
and
\be
T^{BF}_{\g_v,p_f,k_f,I_\t}(g_{f})&=&\prod_{\t}I_{\{j_{\t f}\},\{m_{\t f}\};I_\t}\prod_{f}\Pi^{p_f,k_f}_{j_{\t f}m_{\t f},j_{\t'f}m_{\t'f}}(g_{f})\nonumber
\ee
Recall that we always assume the vertex amplitude  is associated with an integrable spin-network graph, thus is finite after regularization \cite{finite}.

We can project $A_v$ on the new boundary Hilbert space $\ch^L_{\g_v}$, in the same way as the Euclidean case
\be
A_v^L(g_{f})&=&\!\!\sum_{j_f,r_f,i_\t}\prod_f\lt(\frac{\g^2 j_f^2(j_f+1)^2}{(j_f-r_f)^2}+(j_f-r_f)^2\rt)\nonumber \\
&& \lag T^L_{\g_v,j_f,r_f,i_\t}\ ,\ A_v\rag\ T^L_{\g_v,j_f,r_f,i_\t}(g_{\t\t'})
\ee
where $\lag\ ,\ \rag$ is the inner product on the BF boundary Hilbert space. The states
\be
T^L_{\g_v,j_f,r_f,i_\t}(g_{f})=\prod_{\t}i_\t^{A_{\t1}\cdots A_{\t F}}\prod_{(\t,\t')}\Pi^{\frac{\g j_f(j_f+1)}{j_f-r_f},j_f-r_f}_{j_fA_{\t f},j_fA_{\t' f}}(g_{f})
\nonumber
\ee
form an orthogonal basis in $\ch^L_{\g_v}$. By using the orthogonality relation
\be
&&\int_{\slc}\rmd g\ \overline{\Pi^{(p,k)}_{jm,ln}(g)}\ \Pi^{(p',k')}_{j'm',l'n'}(g)=
\nonumber \\
&&\quad\frac{1}{k^2+p^2}\delta^{kk'}\delta(p-p')\delta_{jj'}\delta_{ll'}\delta_{mm'}\delta_{nn'}
\ee
it is straightforward to show that in the $(j_f,r_f,i_\t)$-spin-network representation, the resulting vertex amplitude reads
\be
&&A_v^L\lt(j_f,k_f,i_\t\rt)=\lag T^L_{(\g_v,j_f,k_f,i_\t)}\ ,\ A_v\rag\\ \nonumber
&&\quad=\sum_{I_\t}A_v\Big((\frac{\g j_f(j_f+1)}{j_f-r_f},{j_f-r_f});I_\t\Big)\prod_{\t}f_{I_\t}^{i_\t}
\ee
where
\be
f_{I_\t}^{i_\t}:=\overline{i_\t^{\{A_{\t f}\}}}\ I_{\{j_{f}\},\{A_{\t f}\}I_\t}\lt(\f{\g j_f(j_f+1)}{j_f-r_f},j_f-r_f\rt)
\ee
As expected, the vertex amplitude $A_v^L$ obtained in this manner is divergent, and we need a regularization procedure. To this aim, rewrite the vertex amplitude in terms of spinfoam trace as we did for the Euclidean theory. We define a formal map $I_L^{r_f}$ from $SU(2)$ intertwiners into $\slc$ intertwiners, depending on the quantum numbers $r_f$
\be
I_L^{r_f}(i_\t)_{\{j_f'\},\{A_f'\}}=\int\rmd g\ \prod_{f\subset\t}\Pi^{(\f{\g j_f(j_f+1)}{j_f-r_f},j_f-r_f)}_{j'_{f}A'_{\t f},j_{f}A_{\t f}}\lt(g\rt)\cdot {i_\t^{\{A_{\t f}\}}}
\nonumber
\ee
which gives $A_v^L$ by a spinfoam trace
\be
A^L_v(j_f,r_f,i_\t)=\tr\lt(\bigotimes_{f\in e} I_L^{\{r_f\}}(i_{\t_f})^\dagger\rt)
\ee
To
 regularize the vertex amplitude $A_v^L$ it is sufficient to removing one of the $dg$ integration (which is reduntand) at each vertex. With this, the vertex amplitude $A^L_v$ is finite.

\subsection{Face Amplitude and Partition Function}

It is argued in \cite{face} that the face amplitude of a spinfoam model is determined by three inputs: (a) the choice of the boundary Hilbert space, (b) the requirement that the composition law holds when gluing two complexes $\ck$ and $\ck'$, (c) a particular locality requirement (see \cite{face} for the details of the three assumptions). These requirements are implemented if the partition function has the form \eqref{hr}.  By inserting the vertex amplitudes that we have defined into this expression, we complete the definition of an Euclidean and a Lorentzian model.

Expanding the delta function in representation, we obtain
\be
Z_{E,L}(\ck)&=&\!\!\sum_{j_f,r_f,i_\t}\prod_f
d^{E,L}(j_f,r_f) \ \prod_v A^{E,L}_v(j_f,r_f,i_\t)\nonumber
\ee
where the Euclidean face amplitude is
\be
d^{E}(j_f,r_f)=
\Big[(1+\g)j_{f}+1\Big]\Big[{({1-\g})j_{f}+2r_{f}+1}\Big]
\ee
the Lorentzian one is
\be
d^{L}(j_f,r_f)=\frac{\g^2 j_f^2(j_f+1)^2}{(j_f-r_f)^2}+(j_f-r_f)^2.
\ee
where the dimension factors $A_f^E:=\Big[(1+\g)k_{f}+1\Big]\Big[{({1-\g})k_{f}+2r_{f}+1}\Big]$ and $A^L_f:=\Big[k_f^2+{\g^2 j_f^2(j_f+1)^2}/{k_f^2}\Big]$ are the face amplitudes for the Euclidean and Lorentzian theories. In the Euclidean case, the face amplitudes is different from the one obtained in \cite{face} and coincide with the ones deduced from the BF partition function. In \cite{face} the face amplitude obtained is the dimension of $SU(2)$ unitary irrep i.e. $2j_f+1$. The origin of the difference is the difference in the boundary Hilbert space. The one here, $\ch^E_{\g_v}$ or $\ch^L_{\g_v}$, has additional degree of freedom with respect to the space $L^2(SU(2)^{L})$ of \cite{face}.

\section{The new degree of freedom and relation to quantum GR}

Does the new degree of freedom of the theory defined above, which is captured by the quantum number $r_f$, has a physical interpretation relevant for quantum gravity?  There are some reasons to suspect a negative answer. Let us consider the Euclidean theory for simplicity.

First, we have seen that $r_f$ does not affect the boundary geometry.  We expect all gravitational degrees of freedom to be captured by the geometry. More precisely, in
the classical theory we have the well known (``left area=right area") relation
\be
|\Sigma^+|^2=|\Sigma^-|^2, \label{diag}
\ee which implies
\be
|1-\g|j^+=|1+\g|j^-
\ee
which in turns implies $r_j=0$.  We can still obtain states compatible with GR in the classical limit by demanding that
\be
\lim_{j^\pm\to\infty}\frac{r}{j^-}=0 &\text{for}& 0<\g<1\nonumber\\
\lim_{j^\pm\to\infty}\frac{r}{j^-}=2 &\text{for}& \g>1
\ee
in the large-$j$ asymptotic regime. But this begins to be a bit artificial.

Furthermore, in the classical theory the area of a face can be equally computed in the time gauge as $A_4=\sqrt{(\Sigma_{f})^{IJ}(\Sigma_{f})_{IJ}}$ or as $A_3=\g\sqrt{{\Sigma}_{f}^i{\Sigma}_ f^i}$. Classically the two areas $A_4$ and $A_3$ are equal after the simplicity constraint is imposed, and they indeed equal in the large-$j$ limit after quantization \cite{EPRL}. Let us denote the condition $A_4=A_3$ the \emph{consistency constraint}. If we ask $A_4$ and $A_3$ to be equal as operators in the quantum level on the boundary Hilbert space (as in the case of \cite{EPRL}), then again this fixes $r_f$. The precise value of $r_f$ fixed depends on how the operators corresponding to $A_4$ and $A_3$ are ordered. In this sense the quantum numbers $r_f$ are related to the operator-ordering ambiguities of the consistency constraint. Once an order is chosen, there is no more independent quantum number $r_f$ in the theory. With a suitable ordering, we can fix $r_f=0$

For these consideration, it may be reasonable to suspect that the weak imposition of the simplicity constraints may in fact be too weak to properly define quantum general relativity, in the same sense in which the strong imposition of these constraints in the old Barrett-Crane model was too strong.  There is a simple way out, which is to impose the (non-commuting) simplicity constraints weakly, and the diagonal simplicity constraint (for instance in the form \eqref{diag}) strongly.
With this choice of constraints, properly ordered, we obtain $r_f=0$, precisely the LQG state space in the boundary, and precisely the new models amplitudes.  Finally, the gluing conditions gives the SU(2) face amplitude.  Thus, we recover precisely the quantum gravity theory described for instance in \cite{simple}.

Note that one could also take the point of view that the quantum numbers $r_f$ label different possible definitions of the spin-foam models. In each of these spin-foam models, the boundary Hilbert space solves the simplicity constraint weakly. And for different choices of $r_f$ the boundary Hilbert spaces are isometric to each other.

\section{Conclusion and Outlook}\label{conclusion}

By imposing the simplicity constraints on a quantum BF theory defined on an arbitrary cellular complex, we have obtained a theory which: (1) is well defined both in the Euclidean and the Lorentzian context; (2) generalizes the existing spinfoam model to general 2-cell complexes, along the lines suggested by \cite{kkl}; (3) has boundary state that have a natural interpretation in the semiclassical limit as a polyhedral geometry on the boundary. In particular, we have shown that the KKL extension of the spinfoam formalism still satisfies the simplicity conditions weakly.

The weak simplicity constraint allow a space larger than the one of LQG to emerge. The physical interpretation of the additional degree of freedom is unclear. It can be eliminated by imposing the non-commuting simplicity constraints weakly and the diagonal one strongly.

\section*{Acknowledgments}

The authors are grateful for the fruitful discussions with Eugenio Bianchi and Simone Speziale. Y.D. is supported by CSC scholarship No. 2008604080.


\begin{thebibliography}{20}

\bibitem{simple}
C. Rovelli. Simple model for quantum general relativity from loop quantum gravity.
[arXiv:1010.1939]\\
C. Rovelli. A new look at loop quantum gravity. [arXiv:1004.1780]

\bibitem{sfrevs}
A. Perez. spinfoam models for quantum gravity. Class. Quant. Grav. 20 (2003)
R43-R104. \\
D. Oriti. Spacetime geometry from algebra:
 spin foam models for non-perturbative quantum
gravity. Rep. Prog. Phys. 64 (2001) 1703-1757.\\
J. Baez. Spin foam models. Class. Quant. Grav. 15 (1998) 1827-1858.

\bibitem{BC} J. Barrett and L. Crane. Relativistic spin-networks and quantum gravity. {J. Math. Phys.} {39} 3296\\
J. Barrett and L. Crane. A Lorentzian signature model for quantum general relativity. {Class. Quant. Grav.}  {17} (2000) 3101-3118.

\bibitem{thiemannbook} T. Thiemann. \emph{Modern Canonical Quantum General Relativity} (Cambridge University Press, Cambridge, 2007)

\bibitem{rovellibook} C. Rovelli. \emph{Quantum Gravity} (Cambridge University Press 2004)\\
C. Rovelli and L. Smolin. Loop space representation
for quantum general relativity. Nucl. Phys. B331 (1990) 80

\bibitem{rev}A. Ashtekar and J. Lewandowski. Background independent quantum gravity: A status
report. {Class. Quant. Grav.} {21} (2004) R53.\\
M. Han, W. Huang and Y. Ma. Fundamental structure of loop quantum gravity. Int. J. Mod. Phys.
D16 (2007) 1397-1474 [arXiv:gr-qc/0509064].

\bibitem{perez}K. Noui and A. Perez. Three dimensional loop quantum gravity: physical scalar
product and spin foam models. Class. Quant. Grav. {22} (2006) 1739-1762

\bibitem{links}M. Han and T. Thiemann. On the Relation between Operator Constraint --, Master
Constraint --, Reduced Phase Space --, and Path Integral Quantisation. [arXiv:0911.3428]\\
M. Han and T. Thiemann. On the Relation between Rigging Inner Product and Master Constraint
Direct Integral Decomposition. [arXiv:0911.3431]\\
M. Han. Path Integral for the Master Constraint of Loop Quantum Gravity. [arXiv:0911.3432]\\
J. Engle, M. Han and T. Thiemann. Canonical path-integral measure for Holst and Plebanski
gravity: I. Reduced Phase Space Derivations. [arXiv:0911.3433]\\
M. Han. Canonical path-integral measure for Holst and Plebanski gravity: II. Gauge invariance and
physical inner product. [arXiv:0911.3436]


\bibitem{EPRL}J. Engle, R. Pereira and C. Rovelli. The loop-quantum-gravity vertex-amplitude.
{Phys. Rev. Lett.} {99} (2007) 161301\\
J. Engle, E. Livine, R. Pereira and C. Rovelli. LQG vertex with finite Immirzi parameter. {Nucl.
Phys.} B{799} (2008) 136

\bibitem{FK}L. Freidel and K. Krasnov. New spin foam model for 4d gravity. Class. Quant. Grav.25 (2008) 125018 \\
E. Livine and S. Speziale. A new spinfoam vertex for quantum gravity. {Phys. Rev.} {D76} (2007) 084028\\
E Livine and S Speziale. Consistently solving the simplicity constraints for spinfoam quantum gravity. Europhys. Lett. {81} (2008) 50004




\bibitem{plebanski}J. Plebanski. On the separation of Einsteinian substructures. {J. Math. Phys.}
{18} (1977) 2511-2520. \\
M. P. Reisenberger. Classical Euclidean general relativity from ``left-handed area = righthanded
area''. [arXiv:gr-qc/9804061] \\
R. De Pietri and L. Freidel. so(4) Plebanski action and relativistic spin foam model. {Class. Quant.
Grav.} {16} (1999) 2187-2196.

\bibitem{QSD}T. Thiemann. Quantum spin dynamics. VIII. The master constraint. Class. Quant.Grav. 23 (2006),
2249-2266. [gr-qc/0510011] \\
M. Han and Y. Ma. Master constraint operator in loop quantum gravity. Phys. Lett. B635 (2006),
225-231. [gr-qc/0510014]\\
K. Giesel, T. Thiemann. Algebraic Quantum Gravity (AQG) I,II,III,IV. Class.Quant.Grav.24 (2007) 2465-2588, Class. Quant. Grav. 27 (2010) 175009


\bibitem{DingYou}Y. Ding and C. Rovelli. The volume operator in covariant quantum gravity. \emph{ Class. Quant. Grav.} \textbf{27} (2010)  165003. [arXiv:0911.0543 [gr-qc]]\\
Y. Ding and C. Rovelli. Physical boundary Hilbert space and volume operator in the Lorentzian new spinfoam theory. \emph{ Class. Quant. Grav.} \textbf{27} (2010)  205003. [arXiv:1006.1294[gr-qc]]


\bibitem{barbieri}
A. Barbieri. Quantum tetrahedra and simplicial spin networks. Nucl. Phys. B518 (1998) 714-228

\bibitem{semiclassical}J. W. Barrett, R. J. Dowdall, W. J. Fairbairn, H. Gomes and F. Hellmann. Asymptotic analysis of the EPRL
four-simplex amplitude. J. Math. Phys. 50, 112504 (2009). [arXiv:0902.1170]\\
J. W. Barrett, R. J. Dowdall, W. J. Fairbairn, F. Hellmann and R. Pereira. Lorentzian spin foam amplitudes: graphical calculus and asymptotics. [arXiv:0907.2440]

\bibitem{FC}F. Conrady and L. Freidel. Quantum geometry from phase space reduction. J. Math. Phys. 50 (2009)123510

\bibitem{semiclassical2}
E. Bianchi, E. Magliaro, and C. Perini. LQG propagator from the new spin foams. Nucl.Phys.B822 (2009) 245-269\\
E. Alesci, E. Bianchi, and C. Rovelli. LQG propagator: III. The new vertex. Class. Quant. Grav. 26 (2009) 215001

\bibitem{Bianchi:2010zs}E. Bianchi, C. Rovelli and F. Vidotto.  Towards Spinfoam Cosmology.  Phys.  Rev.  D{82} (2010) 084035.  [arXiv:1003.3483 [gr-qc]]

\bibitem{BCtrouble}J. C. Baez, J. D. Christensen and G. Egan. Asymptotics of 10j symbols. Class.
Quant. Grav. 19 (2002) 6489 \\
L. Freidel and D. Louapre. Asymptotics of 6j and 10j symbols. Class. Quant. Grav. 20 (2003) 1267\\
J. W. Barrett and C. M. Steele. Asymptotics of relativistic spin networks. Class. Quant. Grav. 20
(2003) 1341\\
E. Alesci and C. Rovelli. The complete LQG propagator: I. Difficulties with the Barrett-Crane vertex.
Phys.Rev.D76 (2007) 104012


\bibitem{kkl}W. Kami\'nski, M. Kisielowski and J. Lewandowski. spinfoams for all loop quantum gravity. [arXiv:0909.0939[gr-qc]]\\
W. Kami\'nski, M. Kisielowski and J. Lewandowski. The EPRL intertwiners and corrected partition function. [arXiv:0912.0540[gr-qc]]

\bibitem{Minkowski}H. Minkowski. Allgemeine Lehrs\"atze \"uber die konvexe Polyeder. Nachr. Ges. Wiss. Goettingen (1897) 198-219

\bibitem{poly}E. Bianchi, P. Don\'a, and S, Speziale. Polyhedra in loop quantum gravity. [arXiv:1009.3402v1[gr-qc]]

\bibitem{Ooguri} H. Ooguri. Topological lattice models in four-dimensions. Mod. Phys. Lett. A7 (1992) 2799-2810

\bibitem{propagator}E. Bianchi, L. Modesto, C. Rovelli, and S. Speziale. Graviton propagator in loop quantum gravity. Class. Quant. Grav. 23 (2006) 6989-7028\\
    C. Rovelli. Graviton propagator from background-independent quantum gravity. Phys. Rev. Lett. 97 (2006) 151301

\bibitem{future}Y. Ding and M. Han. Large-j asymptotics of the generalized spinfoam model. [in preparation]

\bibitem{alexei} S. Alexandrov, private communication.

\bibitem{field}W. Greiner and J. Reinhardt. \emph{Field quantization} (Springer, 1996)

\bibitem{string}M. B. Green, J. H. Schwarz, and E. Witten. \emph{Superstring theory: Volume 1 Introduction}. (Cambridge University Press, 1988)

\bibitem{face}E. Bianchi, D. Regoli, and C. Rovelli. Face amplitude of spinfoam quantum gravity. arXiv:1005.0764 [gr-qc]

\bibitem{Magliaro:2010ih}
E. Magliaro, and C. Perini. Local spin foams.  arXiv:1010.5227 [arXiv]

\bibitem{ruhl}
W. Ruhl. \emph{The Lorentz group and harmonic analysis} (W.A. Benjamin, Inc., New York, 1970).\\
I. M. Gel'fand, M. I. Graev, and N. Ya. Vilenkin. \emph{Generalized Functions: Volume 5 Integral Geometry and
Representation Theory} (Academic Press, 1966).

\bibitem{finite}J. Baez and J. Barrett. Integrability for relativistic spin-networks. Class. Quant. Grav. 18 (2001) 4683\\
J. Engle and R. Pereira. Regularization and finiteness of the Lorentzian LQG vertices. Phys. Rev. D79 (2009) 084034

\bibitem{BHNR}E. Buffenoir, M. Henneaux, K. Noui, and Ph. Roche. Hamiltonian analysis of Plebanski theory. {Class. Quant. Grav.} {21} (2004) 5203-5220 [arXiv:gr-qc/0404041]

\bibitem{polyreconstruct}J. B. Lasserre. An analytical expression and an algorithm for the volume of a Convex Polyhedron in Rn. J. Optim. Theor. Appl. 39 (1983) 363-377

\bibitem{gelfand}I. M. Gelfand, N. Ya. Vilenkin. \emph{Generalized functions: Volume 4 Applications of harmonic analysis} (Academic Press, 1964)

\bibitem{gms}I. M. Gel'fand, R. A. Minlos and Z. Ya. Shapiro. {\it Representations of the rotation and Lorentz groups and their applications} (Pergamon Press, 1963), pp. 187-189.

\bibitem{coherent}A. Perelomov. Generalized coherent states and their applications. Springer-Verlag. 1985

\bibitem{Freidel:2009nu}
 L. Freidel, K. Krasnov and E.R. Livine. Holomorphic Factorization for a Quantum Tetrahedron.
 Commun. Math. Phys.  {297}, 45 (2010)
 [arXiv:0905.3627 [hep-th]].

\bibitem{KM}M. Kapovich and J. J. Millson. The symplectic geometry of polygons in Euclidean space. J. Differential Geom. 44, 3 (1996), 479-513.







\end{thebibliography}
\end{document}